\documentclass[journal=jpclcd,manuscript=letter]{achemso}
\usepackage[T1]{fontenc}
\usepackage[latin9]{inputenc}
\usepackage{booktabs}
\usepackage{amsmath}
\usepackage{amssymb}
\usepackage{graphicx}
\PassOptionsToPackage{version=3}{mhchem}
\usepackage{mhchem}

\makeatletter

\usepackage[version=3]{mhchem}

\usepackage{amsfonts}
\usepackage{bm}
\usepackage{setspace}
\usepackage{color}
\usepackage{verbatim}
\usepackage{float}
\usepackage{array}
\usepackage{cleveref}
\usepackage{subfigure}
\usepackage{xr}
\usepackage{threeparttablex}
\usepackage{xcolor}
\usepackage{enumitem}

\title{Self-consistent calculation of localized orbital scaling correction
    for correct electron densities and energy level alignment in
    density functional theory}

\author{Yuncai Mei}
\author{Zehua Chen}
\affiliation{Department of Chemistry, Duke University, Durham,
    North Carolina 27708, USA}
\footnotetext[1]{Y.M. and Z.C. contributed equally to this paper.}

\author{Weitao Yang}
\email{weitao.yang@duke.edu}
\affiliation{Department of Chemistry, Duke University, Durham,
    North Carolina 27708, USA}
\alsoaffiliation{Department of Physics, Duke University, Durham,
    North Carolina 27708, USA}
\date{\today}

\makeatother

\begin{document}
\begin{tocentry}
\includegraphics[width=1\linewidth]{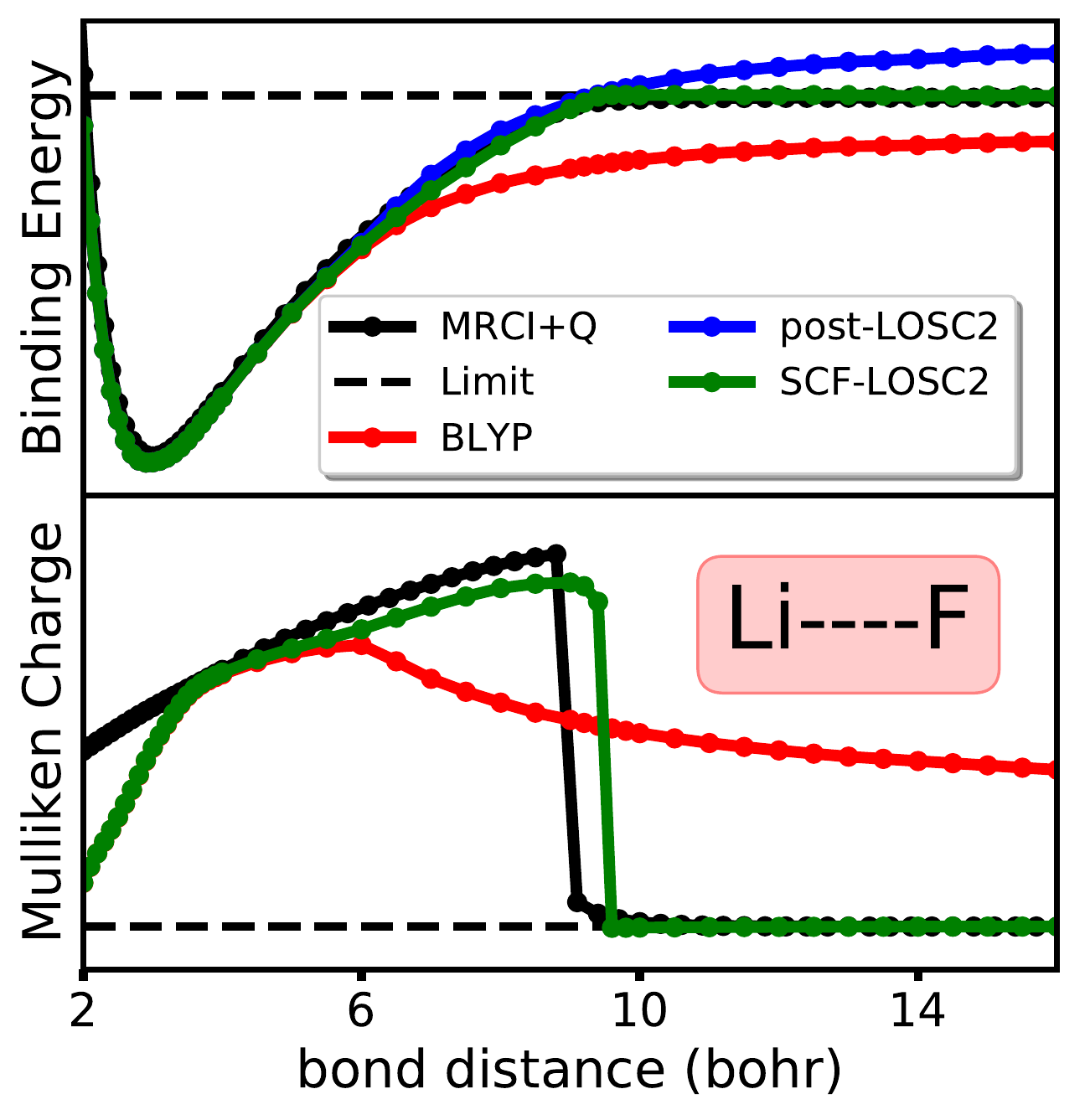} 
\end{tocentry}
\begin{abstract}
The recently developed localized orbital scaling correction (LOSC)
method shows the ability to systematically and size-consistently reduce
the delocalization error existing in conventional density functional
approximations (DFAs).  Applying LOSC to conventional DFAs (LOSC-DFAs)
gives much improvement for the description of related properties,
including band gaps, total energies and photoemission spectra. However,
concern and issue remain: the application of LOSC to DFAs is mainly
through a post self-consistent field (SCF) manner, and few results
from applying LOSC to DFAs with a SCF manner have been reported. The
reason  is  the originally proposed SCF approach for SCF-LOSC calculation
uses an approximate Hamiltonian and encounters convergence problems
easily in practice. In this work, we develop a new SCF approach with
a correct Hamiltonian and achieve reliable SCF-LOSC calculations.
We demonstrate the capability of the new SCF approach for SCF-LOSC 
to correctly describe the electron densities, total energies and energy
level alignment for the molecular dissociation process, while conventional
DFAs or LOSC-DFAs with post-SCF calculations show large errors. This
work demonstrates that the new SCF approach for SCF-LOSC would be
a promising method to study problems for correct electron densities and
energy level alignment in large systems.
\end{abstract}
\maketitle

Density functional theory (DFT)
\cite{hohenbergInhomogeneousElectronGas1964b,
kohnSelfConsistentEquationsIncluding1965b,
parrDensityFunctionalTheoryAtoms1994}
has been widely used to calculate and predict the electronic structure
of molecular systems in practice. The performance of DFT depends on
the quality of applied density functional approximation (DFA)
to the  exchange-correlation energy $E_{{\rm {xc}}}$. Although the
conventional DFAs, such as local density approximation (LDA),
\cite{voskoAccurateSpindependentElectron1980a,
perdewAccurateSimpleAnalytic1992}
general gradient approximations (GGAs)
\cite{beckeDensityfunctionalExchangeenergyApproximation1988a,
leeDevelopmentColleSalvettiCorrelationenergy1988a,
perdewGeneralizedGradientApproximation1996}
and hybrid GGAs,
\cite{stephensInitioCalculationVibrational1994,
adamoReliableDensityFunctional1999,
ernzerhofAssessmentPerdewBurke1999}
are commonly used in practice, they all have the delocalization error
\cite{mori-sanchezLocalizationDelocalizationErrors2008c,
cohenFractionalChargePerspective2008b,
cohenInsightsCurrentLimitations2008b,
cohenChallengesDensityFunctional2012a}
and fail to describe some critical problems.

The delocalization error 
in conventional DFAs \cite{mori-sanchezLocalizationDelocalizationErrors2008c,
liLocalizedOrbitalScaling2018c}
manifests in a size-dependent manner: For small systems, with small
number of atoms and having small physical extent, commonly used DFAs
have good accuracy describing the total energies of systems with integer
number of electrons, but the delocalization error exhibits as the
convex deviation from the Perdew-Parr-Levy-Balduz (PPLB) linear condition
for fractional number of electrons 
\cite{perdewDensityFunctionalTheoryFractional1982a,
yangDegenerateGroundStates2000b,
zhangPerspectiveDensityfunctionalTheory2001};
For large systems, with large number of atoms or having large physical
extent (as near a dissociation limit), commonly used DFAs have small
errors for systems with fractional number of electrons and obey the
fractional charge linearity condition at the bulk limit, but the delocalization
error reveals as the large errors in total energy for systems with
integer number of electrons. Conventional DFAs usually
give much lower total energy for the system with fractional charges,
making the total energy curve convex for small systems.
\cite{yangDegenerateGroundStates2000b,
zhangPerspectiveDensityfunctionalTheory2001}
Such delocalization error (convex deviation) leads conventional DFAs
to produce large error for the energy derivatives with respect to
the electron number, i.e. the chemical potentials,
\cite{cohenFractionalChargePerspective2008b,
yangDerivativeDiscontinuityBandgap2012a}
leading to underestimation of the exact  ionization potentials (IPs)
from the HOMO (highest occupied molecular orbital) energies and the
overestimation of electron
affinities (EAs) from the LUMO (lowest unoccupied molecule orbital)
energies. Delocalization
error would produce too delocalized electron density, as the error
falsely lowers the total energy of the system.
\cite{cohenInsightsCurrentLimitations2008b,
mori-sanchezLocalizationDelocalizationErrors2008c,
cohenChallengesDensityFunctional2012a}
Examples reflecting this issue are the wrong dissociation limits of
molecules
\cite{dutoiSelfinteractionErrorLocal2006,
mori-sanchezManyelectronSelfinteractionError2006,
ruzsinszkySpuriousFractionalCharge2006,
vydrovTestsFunctionalsSystems2007,
zhengDelocalizationErrorDensityfunctional2012}
and too low reaction barriers \cite{cohenChallengesDensityFunctional2012a}
from conventional DFAs. For charge transfer complexes, delocalization
error leads to overestimate of charge transfer 
and the binding energies \cite{cohenInsightsCurrentLimitations2008b}. 

In material interfaces and defects, delocalization error can lead
to incorrect charge transfer across the interfaces, and significant
error in energy level alignment 
\cite{Flores098658,Souza13165112,Pacchioni1580,Egger152448}.
Interfaces and energy level alignment play important roles in many
technological applications: they strongly influence the charge extraction
and transport in solar cell devices \cite{Wang181800260}, and catalysis
in semiconductor \cite{hegnerLevelAlignmentDescriptor2017}. Thus
it remains an important challenge to describe the correct energy level
alignment for interfaces with DFAs.

To reduce the delocalization error, there has been much effort devoted
to develop improved functional approximations. These include the development
of long-range corrected functionals \cite{
Savin95327,gillCoulombattenuatedExchangeEnergy1996,
leiningerCombiningLongrangeConfiguration1997a,
iikuraLongrangeCorrectionScheme2001,
Toulouse04062505,
yanaiNewHybridExchange2004a,
baerDensityFunctionalTheory2005,
vydrovAssessmentLongrangeCorrected2006,
vydrovImportanceShortrangeLongrange2006,
cohenDevelopmentExchangecorrelationFunctionals2007,
Chai08084106,
baerTunedRangeSeparatedHybrids2010}
and double hybrid functionals
\cite{zhaoDoublyHybridMeta2004,
grimmeSemiempiricalHybridDensity2006,
chaiLongrangeCorrectedDoublehybrid2009,
zhangDoublyHybridDensity2009,
suFractionalChargeBehavior2014}.
These methods have been shown to reduce delocalization error. But
challenges still remain for a systematic correction across system
types, sizes and scales.

Recently, our group developed the localized orbital scaling correction
(LOSC) method, which imposes the PPLB condition by utilizing orbitalets
(a set of molecular orbitals localized on both physical and energy
space) to the associated parent DFAs to reduce the delocalization error.
\cite{liLocalizedOrbitalScaling2018c,suPreservingSymmetryDegeneracy2020a}
Benefitting from the novel features of orbitalets that dynamically
switch between the canonical orbitals (COs) and localized orbitals
(LOs), correction from LOSC can be flexibly and automatically applied
on the global or local region of the system. Thus, LOSC shows the
ability to reduce the delocalization error in a systematic and size-consistent
way. LOSC has been shown to improve greatly the prediction of quasiparticle
energies -- ionization energies, electron affinities, and also ionized
excited state energies for atoms, small molecules and very large systems,
all from the eigenvalues of the generalized Kohn-Sham Hamiltonian.
\cite{liLocalizedOrbitalScaling2018c,
suPreservingSymmetryDegeneracy2020a,
meiApproximatingQuasiparticleExcitation2019b}
The LOSC prediction accuracy is similar or better than many-electron
Green's function method of GW.
\cite{meiApproximatingQuasiparticleExcitation2019b}

In the original LOSC paper \cite{liLocalizedOrbitalScaling2018c},
two approaches for applying LOSC to the parent DFAs have been discussed.
One way is in the post self-consistent field manner (post-LOSC), in
which the converged electron density from the parent DFA ($\rho_{s}^{{\rm {DFA}}}$)
is directly used to evaluate the energy correction $\Delta E^{{\rm {LOSC}}}$
from LOSC. 
The other way is in a self-consistent field manner (SCF-LOSC), in
which the LOSC effective Hamiltonian $\Delta h^{{\rm {LOSC}}}=\frac{\delta\Delta E^{{\rm {LOSC}}}}{\delta\rho_{s}}$
is introduced to the DFA Hamiltonian $h_{s}^{{\rm {DFA}}}$. After
solving the KS-equations with updated Hamiltonian $h_{s}=h_{s}^{{\rm {DFA}}}+\Delta h^{{\rm {LOSC}}}$
self-consistently, the converged electron density from LOSC-DFA ($\rho_{s}^{{\rm {LOSC-DFA}}}$)
is obtained and the correction $\Delta E^{{\rm {LOSC}}}$ is therefore
evaluated based on $\rho_{s}^{{\rm {LOSC-DFA}}}$. In practice, the
originally proposed SCF approach \cite{liLocalizedOrbitalScaling2018c}
for the SCF-LOSC calculation turns to encounter convergence problems
easily, especially for the calculations of large molecules, because
only an approximate form of the LOSC effective Hamiltonian has been
developed and used. Therefore, only the performance of post-LOSC has
been well investigated along the development of LOSC. Although the
post-LOSC has been demonstrated to show much improvement to the description
of band gaps, total energies and photoemission spectra,
\cite{liLocalizedOrbitalScaling2018c,
meiApproximatingQuasiparticleExcitation2019b,
suPreservingSymmetryDegeneracy2020a}
the development of reliable SCF-LOSC approach is still necessary because
of the following. First, besides the energetic properties (total energies
and orbital energies), the electron density of a molecular system
is also an important property, since it closely relates to the molecule's
geometry, chemical bonding and reaction reactivities. The conventional
DFAs suffering from the delocalization error can produce much
delocalized electron density and underestimate total energy in many
cases. \cite{dutoiSelfinteractionErrorLocal2006,
mori-sanchezManyelectronSelfinteractionError2006,
ruzsinszkySpuriousFractionalCharge2006,
vydrovTestsFunctionalsSystems2007,
zhengDelocalizationErrorDensityfunctional2012}
Applying post-LOSC in these cases is not sufficient because it only
improves the energies for the parent DFA but leaves the significant
error in electron density unchanged. Second, post-LOSC being an reasonable
approximation to the SCF-LOSC is only valid at the condition that
$\rho_{s}^{{\rm {DFA}}}$ is close to $\rho_{s}^{{\rm {LOSC-DFA}}}$.
When $\rho_{s}^{{\rm {DFA}}}$ differs much to $\rho^{{\rm {LOSC-DFA}}}$
as the delocalization error from the parent DFA becomes significant,
the post-LOSC may not provide reliable results due to the lack of
self-consistency, and applying SCF-LOSC calculation becomes necessary.

In this work, we present a new and robust SCF approach to achieve
reliable SCF-LOSC calculations. To start, we briefly review the methodology
of LOSC. The energy correction from LOSC, $\Delta E^{{\rm {LOSC}}}$,
is constructed from a set of orbitalets (LOs,
$\{\phi_{i}\}$) those are transformed from the canonical orbitals
(COs, $\{\psi_{i}\}$) by a unitary transformation. $\Delta E^{{\rm {LOSC}}}$
is expressed as 
\begin{equation}
    \Delta E^{{\rm {LOSC}}}=\sum_{ij}\frac{1}{2}\kappa_{ij}\lambda_{ij}(\delta_{ij}-\lambda_{ij}),
    \label{eq:E_corr}
\end{equation}
in which $\kappa$ is the curvature matrix and $\lambda$ is the local
occupation matrix. The curvature matrix shows different expressions
in different versions of LOSC. We call the original version of LOSC
as LOSC1 \cite{liLocalizedOrbitalScaling2018c} and the later version
as LOSC2 \cite{suPreservingSymmetryDegeneracy2020a} in the following
text. No matter in which version of LOSC, the curvature matrix is
completely and explicitly determined by LOs. For example, the curvature
matrix from LOSC1 is defined as 
\begin{equation}
    \kappa_{ij}=\int\frac{\rho_{i}(\mathbf{r})\rho_{j}(\mathbf{r'})}
        {|\mathbf{r}-\mathbf{r'}|}\text{d}\mathbf{r}\text{d}\mathbf{r'}
        - \frac{2\tau C_{x}}{3} \int[\rho_{i}(\mathbf{r})]^{\frac{2}{3}}
        [\rho_{j}(\mathbf{r})]^{\frac{2}{3}}\text{d}\mathbf{r},
    \label{eq:kappa}
\end{equation}
where $\rho_{i}(\mathbf{r})$ is the local orbital's density and defined
as $\rho_{i}(\mathbf{r})=|\phi_{i}(\mathbf{r})|^{2}$ . The local
occupation matrix $\lambda$ is completely and explicitly determined
by LOs and the KS density operator $\rho_{s}$, and it is expressed
as 
\begin{equation}
    \lambda_{ij}=\langle\phi_{i}|\rho_{s}|\phi_{j}\rangle.
    \label{eq:localocc}
\end{equation}

\begin{figure}[htbp]
    \centering \includegraphics[width=0.6\linewidth]{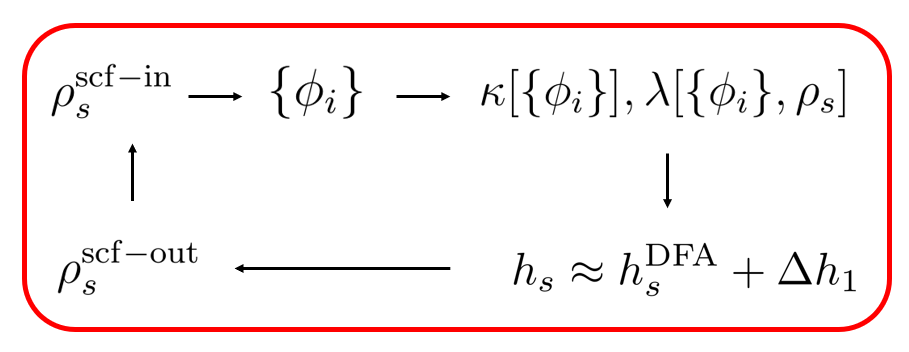}
    \caption{The original SCF procedure with the approximate LOSC effective Hamiltonian.}
    \label{fig:old_scf} 
\end{figure}

Figure \ref{fig:old_scf} shows the working flow of the original SCF
procedure for SCF-LOSC.\cite{liLocalizedOrbitalScaling2018c} 
In particular, one should notice that both LOs and electron density
are updating in every SCF cycle. The dependence of LOs on the electron
density in each SCF cycle introduces complexities to derive the exact
LOSC effective Hamiltonian.
Specifically, according to the original SCF procedure and Eqs. \ref{eq:E_corr}
- \ref{eq:localocc}, the exact LOSC effective Hamiltonian $\Delta h^{{\rm {LOSC}}}$
for the original SCF approach is given by 
\begin{equation}
    \Delta h^{{\rm {LOSC}}}=\frac{\delta\Delta E[\{\phi_{i}\},\rho_{s}]}{\delta\rho_{s}}
        =\frac{\delta\Delta E}{\delta\rho_{s}}\Big|_{\{\phi_{i}\}}
            + \sum_{i}\frac{\delta\Delta E}{\delta\phi_{i}}\Big|_{\rho_{s}}\frac{\delta\phi_{i}}{\delta\rho_{s}}
            + \sum_{i}\frac{\delta\Delta E}{\delta\phi_{i}^{*}}\Big|_{\rho_{s}}\frac{\delta\phi_{i}^{*}}{\delta\rho_{s}}.
    \label{eq:losc_H}
\end{equation}
The first term (denoted as $\Delta h_{1}$) on the r.h.s. of Eq. \ref{eq:losc_H}
is the explicit contribution with the LOs fixed, and the last two
terms (denoted as $\Delta h_{2}$) are the implicit contribution from
the relaxation of LOs with the updating of electron density. Due to
the difficulty of evaluating $\Delta h_{2}$, the $\Delta h_{2}$
term was ignored in practical calculations. As shown in Figure \ref{fig:old_scf},
the LOSC effective Hamiltonian is approximated with only the $\Delta h_{1}$
term, 
\begin{equation}
    \Delta h^{{\rm {LOSC}}}\approx\Delta h_{1}=
        \sum_{i}\kappa_{ii}(\frac{1}{2}-\lambda_{ii})|\phi_{i}\rangle\langle\phi_{i}|
        - \sum_{i\neq j}\kappa_{ij}\lambda_{ij}|\phi_{i}\rangle\langle\phi_{j}|.
    \label{eq:losc_h1}
\end{equation}
However, such approximate LOSC effective Hamiltonian is not robust
in practice, because we find it is easy to cause SCF convergence problem.
To solve the convergence problem, one straightforward solution is
to derive the $\Delta h_{2}$ term to use the exact LOSC effective
Hamiltonian. However, this would be complicated and difficult to
achieve.

In this paper, we develop an alternative solution to the problem.
The idea is to define a new SCF procedure 
with the removal of LOs' dependence on the electron density in
each SCF cycle.
The key step is to redefine the
LOSC localization procedure, in another word, the localization cost functional.
Taking the LOSC2 localization cost functional as an example, the cost
functional $F$ takes the following form,
\begin{align}
    F(\rho_s, \{\psi_i\}, U_{pi}) &= (1 - \gamma) \sum_p \Big(\langle \mathbf{r}^2\rangle_p
        - \langle \mathbf{r} \rangle_p^2 \Big) + \gamma C
        \sum_p \Big( \langle h[\rho_s]^2]\rangle_p
        - \langle h[\rho_s] \rangle_p^2    
        \Big), \\
    \langle X \rangle_p &= \langle \phi_p | X | \phi_p \rangle, \quad X = \mathbf{r},
    \mathbf{r}^2, h, h^2, \\
    \phi_p &= \sum_{i} U_{pi} \psi_i,
\end{align}
in which $U$ is the unitary transformation matrix, $h$ is the
one-electron Hamiltonian of the parent DFA evaluated
at the electron density $\rho_s$ at each SCF cycle, and $\{\psi_i\}$ are
the corresponding COs.
Instead of defining the LOSC localization cost functional
that is dependent on the electron density and COs in each SCF cycle as shown above,
we can just use the electron density $\rho_s^{\rm{DFA}}$ and the
corresponding COs $\{\psi_i^{\rm{DFA}}\}$ from a converged DFA calculation
in the cost functional; thus, we obtain a set of predetermined LOs
$\{\phi_{i}^0\}$ in advance of the SCF-LOSC calculation. Applying
this strategy to LOSC2 as an example, the modified localization cost function is
expressed as
\begin{align}
    F(\rho_s^{\rm{DFA}}, \{\psi_i^{\rm{DFA}}\}, U_{pi}) &= (1 - \gamma) \sum_p \Big(\langle \mathbf{r}^2\rangle_p
        - \langle \mathbf{r} \rangle_p^2 \Big) + \gamma C
        \sum_p \Big( \langle h[\rho_s^{\rm{DFA}}]^2]\rangle_p
        - \langle h[\rho_s^{\rm{DFA}}] \rangle_p^2    
        \Big), \\
    \langle X \rangle_p &= \langle \phi_p^0 | X | \phi_p^0 \rangle, \quad X = \mathbf{r},
    \mathbf{r}^2, h, h^2, \\
    \phi_p^0 &= \sum_{i} U_{pi} \psi_i^{\rm{DFA}},
\end{align}
in which $h$ is now evaluated at the $\rho_s^{\rm{DFA}}$.
Clearly, the LOs in this new SCF approach do not depend on the electron
density of each SCF cycle. We keep the same set of LOs $\{\phi_i^0\}$
during the SCF procedure. This treatment of
LOs in the new approach makes the $\Delta h_{2}$ term
in Eq. \ref{eq:losc_H} vanish
and gives the exact LOSC Hamiltonian only with the $\Delta h_{1}$
term immediately; that is,
\begin{equation}
    \Delta h^{{\rm {LOSC}}}=\frac{\delta\Delta E[\{\phi_{i}^{0}\},\rho_{s}]}{\delta\rho_{s}}=\sum_{i}\kappa_{ii}(\frac{1}{2}-\lambda_{ii})|\phi_{i}^{0}\rangle\langle\phi_{i}^{0}|
        - \sum_{i\neq j}\kappa_{ij}\lambda_{ij}|\phi_{i}^{0}\rangle\langle\phi_{j}^{0}|.
    \label{eq:losc_H_new}
\end{equation}
The LOSC curvature
is determined by the set of $\{\phi_{i}^{0}\}$, and it only needs
to be evaluated once. The local occupation matrix is evaluated based
on $\{\phi_{i}^{0}\}$ and electron density during the SCF procedure;
that is, $\lambda_{ij}=\langle\phi_{i}^{0}|\rho_{s}|\phi_{j}^{0}\rangle$.

\begin{figure}[htbp]
    \centering \includegraphics[width=0.6\linewidth]{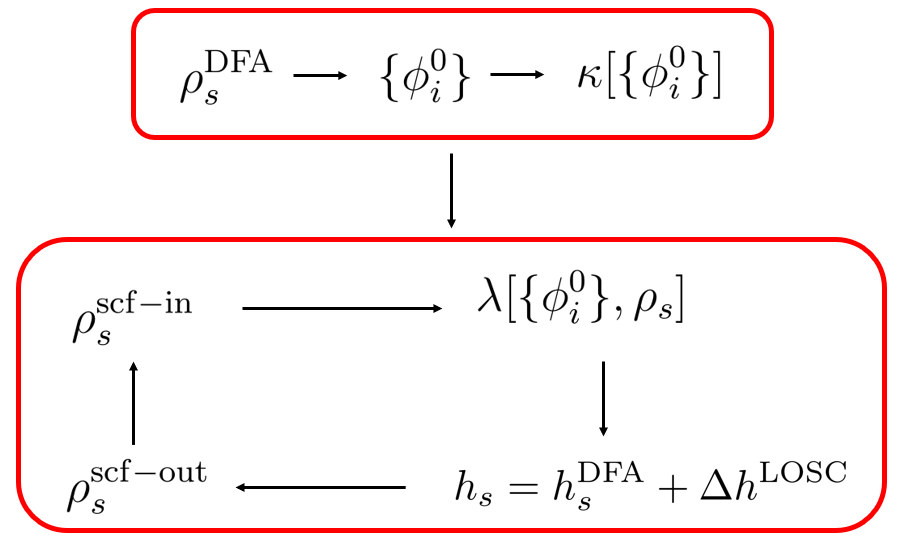}
    \caption{The new SCF procedure with the exact LOSC  Hamiltonian.}
    \label{fig:new_scf} 
\end{figure}

The working flow of the new SCF procedure for SCF-LOSC calculation
is demonstrated in Figure \ref{fig:new_scf}. It involves the following
steps: (1) carry out the SCF convergence from the parent DFA to get
the converged electron density $\rho_{s}^{{\rm {DFA}}}$ and
COs $\{\psi_i^{\rm{DFA}}\}$; (2) apply
the LOSC localization procedure to generate $\{\phi_{i}^{0}\}$ based
on $\rho_{s}^{{\rm {DFA}}}$ and $\{\psi_i^{\rm{DFA}}\}$; (3) construct and store the curvature
matrix evaluated from $\{\phi_{i}^{0}\}$; (4) use the $\rho_{s}^{{\rm {DFA}}}$
as the initial guess to start the SCF-LOSC calculation associated
with parent DFA; (5) use the density $\rho_{s}$ from current SCF
cycle to construct the DFA Hamiltonian $h_{s}^{{\rm {DFA}}}$; (6)
evaluate local occupation matrix based on $\rho_{s}$ and $\{\phi_{i}^{0}\}$,
and construct the exact LOSC effective Hamiltonian $\Delta h^{{\rm {LOSC}}}$
via Eq. \ref{eq:losc_H_new}; (7) apply $\Delta h^{{\rm {LOSC}}}$
to $h_{s}^{{\rm {DFA}}}$ and update COs and $\rho_{s}$; (8) check
the convergence and go back to step (5) if it is not converged.

In the new SCF-LOSC approach, since the LOSC curvature matrix only
needs to be evaluated once and updating the local occupation, $\lambda_{ij}=\langle\phi_{i}^{0}|\rho_{s}|\phi_{j}^{0}\rangle,$
is simple, the computational cost for the new SCF-LOSC approach is
only about two times that of the conventional DFA SCF calculation.
Specifically, one is  the generation of LOs from a one-time conventional
DFA SCF calculation, and the other is  the SCF-LOSC calculation with
the corrected KS Hamiltonian, with fixed LOs and LOSC curvature.

Comparing the new SCF approach with the original one, the SCF solution
from the new SCF approach may be different from the original SCF-LOSC
solution. This is because, at the SCF solution point, the LOs used
to evaluate the total energy in the new SCF-LOSC approach are always
obtained from $\rho_{s}^{{\rm {DFA}}}$, rather than $\rho_{s}^{{\rm {LOSC-DFA}}}$.
The significance of this difference needs to be verified with numerical
results. If the relaxation of LOs, like in the original SCF-LOSC,
turns to be necessary, we can apply an additional layer of SCF cycle
on top of the new SCF-LOSC procedure in order to update the LOs. This
two-layer SCF method is noted as macro-SCF-LOSC. Detailed procedure
for the macro-SCF-LOSC is described in the supporting information.
Ideally, because the macro-SCF-LOSC optimizes both electron density
and LOs, the macro-SCF-LOSC should yield the same results as the original
SCF-LOSC approach. From the numerical results shown in the following
main text, we find that the macro-SCF-LOSC is not necessary in practice,
because the new SCF approach without the macro iteration is already
able to provide reliable and excellent results.

\begin{figure}[htbp]
    \centering
    \subfigure[SCF-LOSC2]{\includegraphics[width=0.48\linewidth]{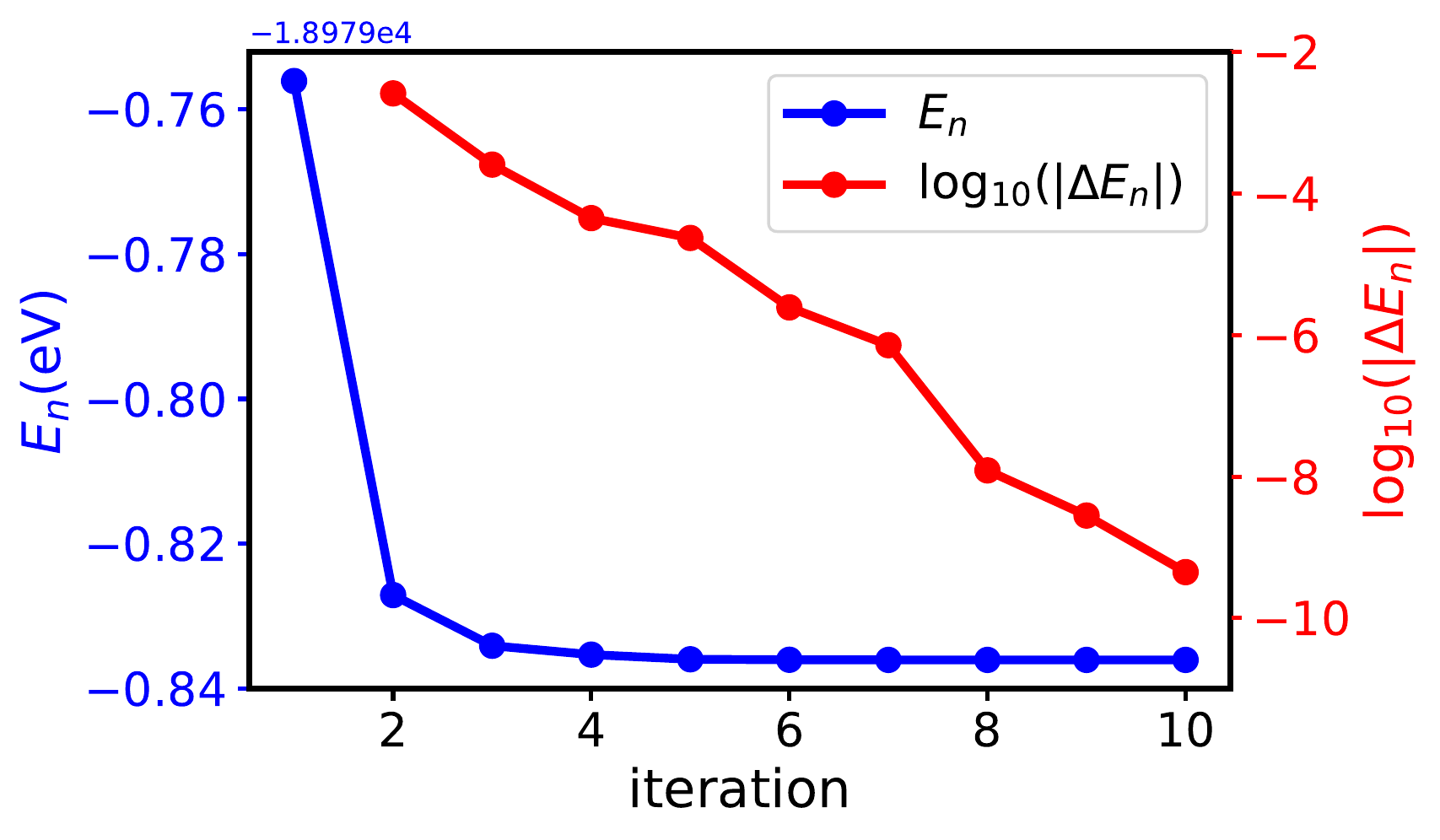}}
    \subfigure[Approx-SCF-LOSC2]{\includegraphics[width=0.48\linewidth]{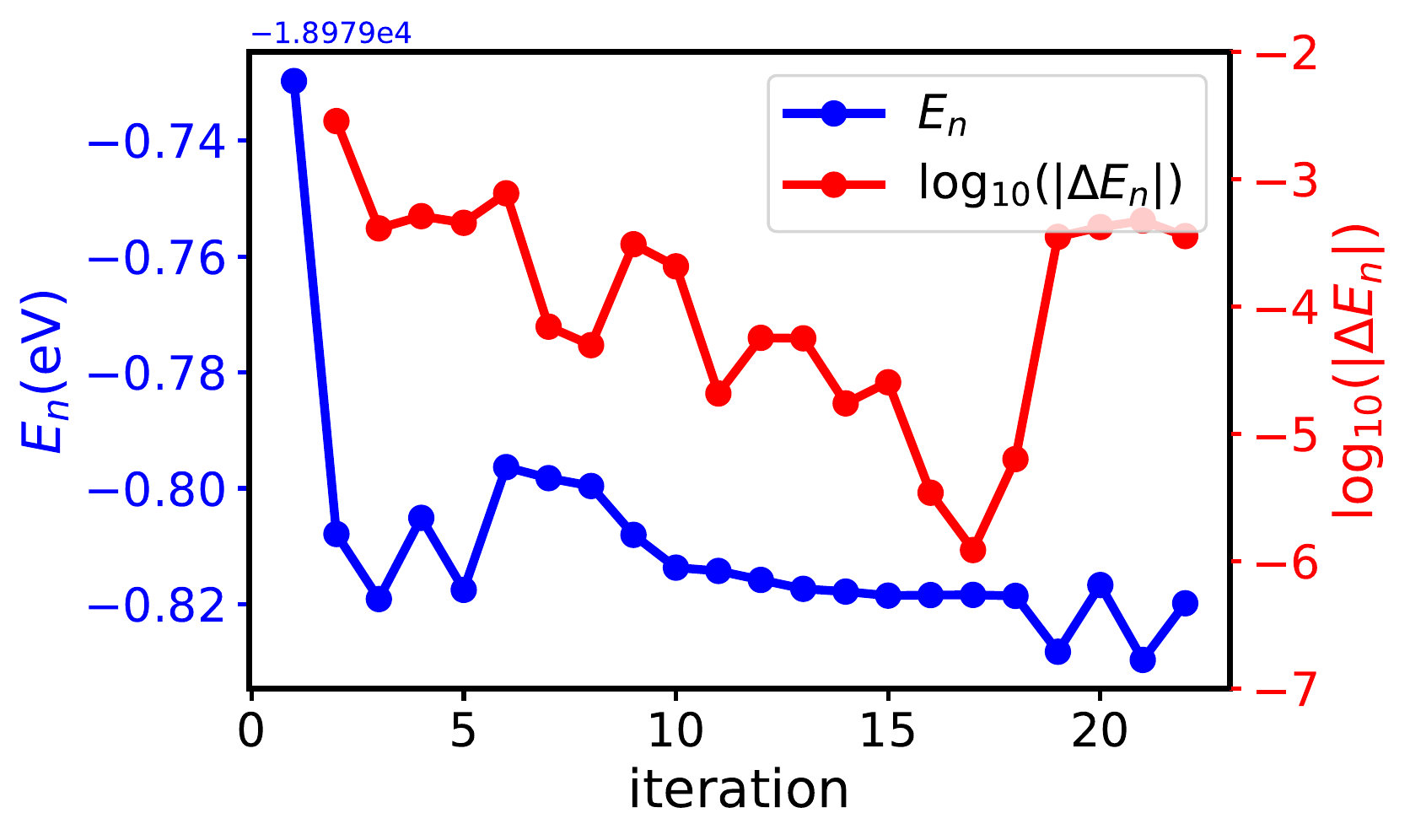}}
    \caption{Comparison of SCF performance between (a) SCF-LOSC2 and (b) the original
        SCF-LOSC2 with approximate Hamiltonian (Approx-SCF-LOSC2) for polyacetylene
        (n = 9). $E_{n}$ is the total energy at $n$-th iteration step and
        $\Delta E_{n}=E_{n}-E_{n-1}$. BLYP is used as the parent functional.
        6-31g{*} is used as the basis set. aug-cc-pVTZ-RIFIT is used as the
        density fitting basis in LOSC2 curvature matrix construction.}
    \label{fig:scf} 
\end{figure}

Now, we check the performance of the new SCF approach for SCF-LOSC
calculations. In the following text, the results from SCF-LOSC refer
to the new SCF approach, if not specified otherwise. We first study
the SCF convergence by testing a long organic molecule, the polyacetylene
with 9 units of monomer (\ce{{[}C2H2{]}9}). As shown in Figure \ref{fig:scf},
the original SCF approach with the approximate LOSC effective Hamiltonian
shows oscillation for total energy along the iterative steps, and
it can not reach to the convergence after about 20 steps. The new
SCF procedure with the exact LOSC effective Hamiltonian converges
smoothly and quickly.

Next, we study the effect of initial guess in the new SCF approach.
Note that in step (4) of the new SCF procedure, we use
$\rho_{s}^{{\rm {DFA}}}$ as the SCF initial guess. This is necessary
because the geometric orientation of LOs are fixed in the SCF process.
Using $\rho_{s}^{{\rm {DFA}}}$ as the initial guess makes sure the
orientation of the LOs agree with the one of initial density and initial
COs at the beginning. If the used initial electron density has a different
orientation than that of the LOs, it will artificially produce fractional local
occupation numbers at the initial step, which may lead the SCF calculation
converge to unphysical state with wrong energies. To support the discussion,
Table \ref{tab:init_guess} shows results for the test on F atom,
with a partially filled p shell. For such small system, the LOs from
LOSC localization are just the converged COs from the parent DFA.
In other words, the localization is not operative and orbitaltes are
just the COs for small systems. As shown in Table \ref{tab:init_guess},
if we use the non-rotated $\rho_{s}^{{\rm {DFA}}}$ as the initial
guess, the local occupation number will be exact integer (either 1
or 0), and LOSC (both LOSC1 and LOSC2) gives zero correction to the
parent DFA. However, if we use the rotated $\rho_{s}^{{\rm {DFA}}}$
as the initial guess, which has a set of rotated p orbitals, the fractional
local occupation number will be generated artificially and leads non-zero
correction to the parent DFA. In the case of LOSC1, we can see such
artificial correction is even negative and leads to a unphysical state
with energy even lower than DFA. In the case of LOSC2, the artificial
energy correction is much smaller. This is because LOSC2 preserves
the symmetry much more than LOSC1 \cite{suPreservingSymmetryDegeneracy2020a},
making these artificial energy corrections from each fractional local
occupation number almost canceled with each other. As a whole, to
avoid issues introduced by the orientation of the initial
guess, we use the $\rho_{s}^{{\rm {DFA}}}$ as the initial
guess for the new SCF approach. Note, the choices of LO orientation
will not be an issue at all, if the macro-SCF-LOSC is used, because
the LO will be generated from $\rho_{s}^{{\rm {LOSC-DFA}}}$ in the
macro iterations \cite{SI}.

\begin{table}
    \caption{Testing on F atom for the effect of the orientation of initial guess
        to the new SCF-LOSC approach. BLYP is used as the parent functional.
        The converged densities from BLYP (rotated/non-rotated) are used as
        the initial guess. cc-pVTZ is used the basis set. aug-cc-pVTZ-RIFIT
        is used as the density fitting basis set in LOSC curvature matrix
        construction. Grid type is (99, 590).}
    \label{tab:init_guess}
    \begin{tabular}{@{}lllll@{}}
    \toprule
    Method               & E\footnotemark[3]              &
    $\Delta$E\footnotemark[4]& E\_corr(initial)\footnotemark[5]  & E\_corr(SCF)\footnotemark[6]
     \\ \midrule
    BLYP                 & -99.7522856431 &           &                  &              \\
    \midrule
    SCF-LOSC1: non-rotated\footnotemark[1]   & -99.7522856430 & 6.31E-11  & 1.10E-14         & 1.30E-14     \\
    SCF-LOSC1: rotated\footnotemark[2] & -99.7794002845 & -2.71E-02 & -2.68E-02        & -2.71E-02    \\
    \midrule
    SCF-LOSC2: non-rotated\footnotemark[1]  & -99.7522856430 & 6.31E-11  & 1.00E-14         & 8.00E-15     \\
    SCF-LOSC2: rotated\footnotemark[2]& -99.7522340746 & 5.16E-05  & 5.27E-05         & 5.05E-05     \\ \bottomrule
    \end{tabular}
    \\
    \footnotemark[1]{The orientation of initial electron density matches
    to the one of LOs.}
    \footnotemark[2]{The orientation of initial electron density does not match
    to the one of LOs.}
    \footnotemark[3]{Total energy.}
    \footnotemark[4]{Total energy difference between the SCF-LOSC and BLYP.}
    \footnotemark[5]{LOSC energy correction at the first SCF cycle.}
    \footnotemark[6]{LOSC energy correction at the last SCF cycle.}
\end{table}

With the smooth convergence from the new SCF procedure,
we investigate the new SCF-LOSC with the same test sets
used in the development of LOSC to test the performance for the atomization
energies, reaction barriers, first ionization potentials (IPs)
and electron affinities (EAs). Detailed results are documented in
the supporting information. In general, the new SCF-LOSC can be conducted
easily in these test sets. For the test sets related to atomization
energies and reaction barriers, both SCF-LOSC and post-LOSC currently
preserve the performance as DFAs.
For example, the mean absolute error (MAE) of G2-1 test set for
atomization energies is 4.94 kcal/mol for BLYP, and 4.93 kcal/mol
for both post-LOSC2-BLYP and SCF-LOSC2-BLYP.
The MAE of HTBH38 test case for the reaction barriers is 7.67 kcal/mol
for BLYP, 7.49 kcal/mol for post-LOSC2-BLYP, and 7.54 kcal/mol for
SCF-LOSC2-BLYP.
Such performance is expected because most of cases in
these tests are with small molecular sizes and large orbital energy
gaps between HOMO and LUMO, which makes orbitalets equal to the
COs of the parent DFA and the local occupation matrix being diagonal
with integer numbers (1 for occupied space and 0 for virtual space).
According to energy correction from LOSC shown in Eq. \ref{eq:E_corr},
these integer local occupation numbers give zero correction to the
total energies. Re-tuning the parameters in the LOSC localization
cost function to obtain more balanced localization between the physical
space and energy space should provide better performance. \cite{suPreservingSymmetryDegeneracy2020a}
However, this task is beyond the scope of current work and will be
studied in the future. For the test sets related to IPs and EAs, new
SCF-LOSC produces much improvement compared to the parent DFA, and
its performance is very similar to the one from post-LOSC for the
tested cases.
For example, the MAE of IP test set shown in the supporting
information is 4.50 eV for BLYP, and 0.62 eV for both post-LOSC2-BLYP
and SCF-LOSC2-BLYP.

In the following, we mainly focus on presenting the results that
can be significantly different, all related to electron density and
energy levels (quasiparticle energies)
associated with the molecular binding/dissociation
processes. We first investigate the dissociation of three
diatomic molecules (LiF, LiH and HF). B3LYP \cite{beckeDensityFunctionalThermochemistry1993a,leeDevelopmentColleSalvettiCorrelationenergy1988a,beckeDensityfunctionalExchangeenergyApproximation1988a}
is applied as the parent functional, because the SCF convergence from
B3LYP can be easily reached at long bond distance for these molecules.
Similar GGA calculations would show even larger delocalization error 
with charge density, but was not obtained because of the SCF failure
for large bond lengths. Results from multireference configuration
interaction method with the Davidson correction (MRCI+Q) \cite{knowlesEfficientMethodEvaluation1988,langhoffConfigurationInteractionCalculations1974,wernerEfficientInternallyContracted1988}
are used as the reference and compared to the results from B3LYP,
post-LOSC-B3LYP and SCF-LOSC-B3LYP.
To study the description of electron density, we look at the Mulliken
charges from Mulliken population analysis 
\cite{mullikenElectronicPopulationAnalysis1955}
and the relative total energy of the molecule to its dissociation
limit along the dissociation process. 

\begin{figure}[htbp]
    \centering
    \subfigure[Mulliken charge]{\includegraphics[width=0.48\linewidth]{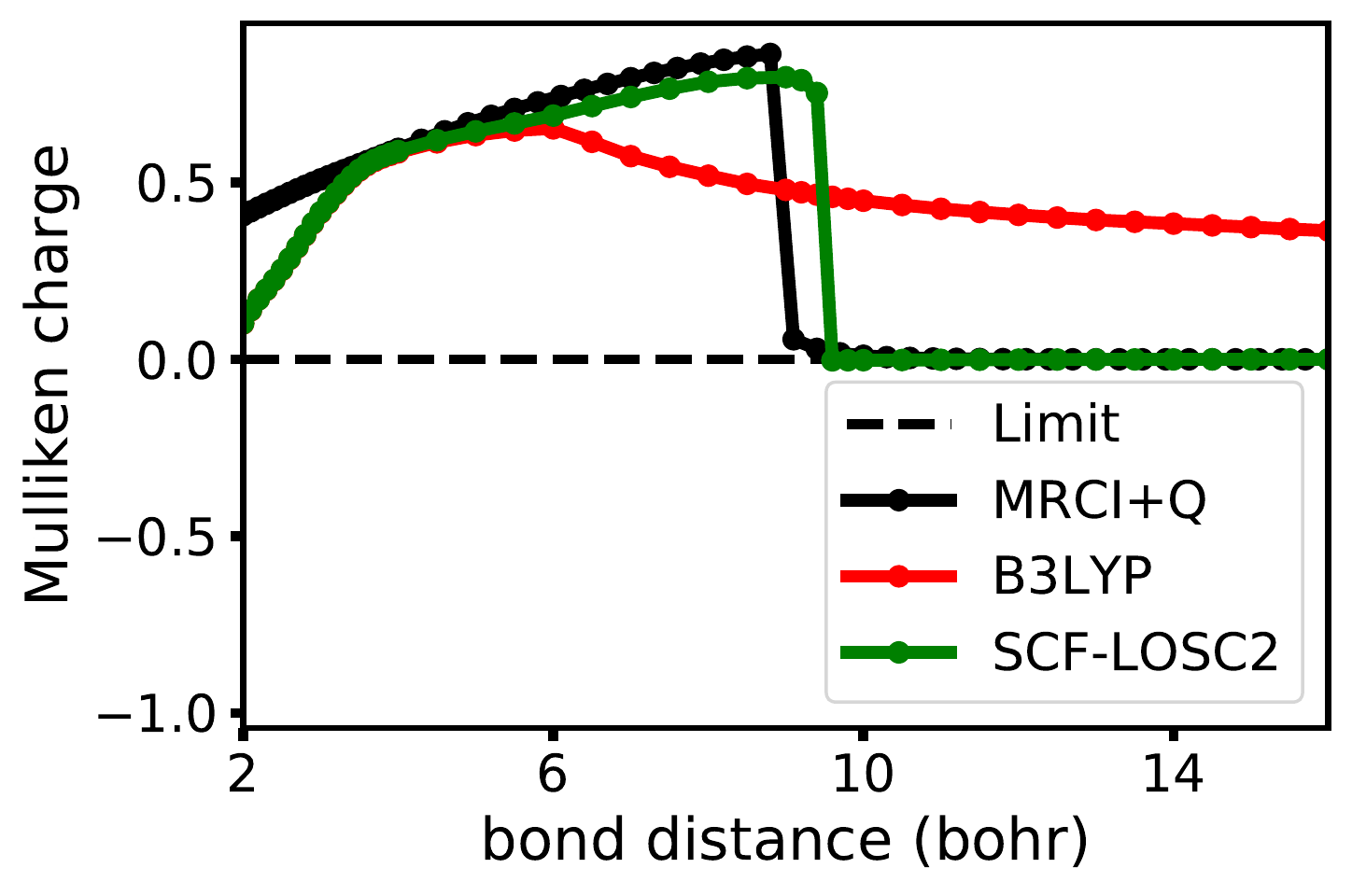}}
    \subfigure[Relative total energy]{\includegraphics[width=0.48\linewidth]{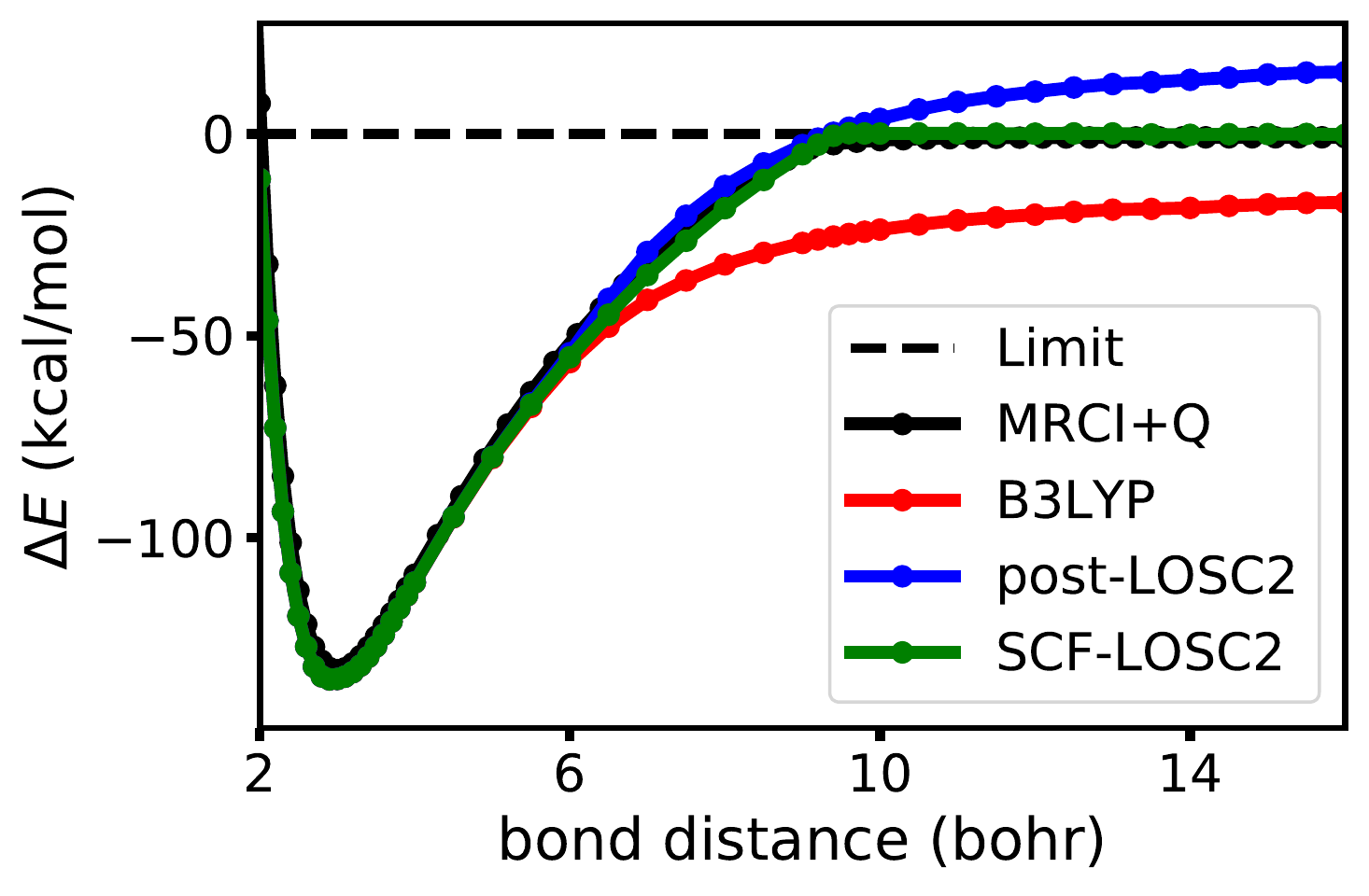}}
    \caption{Dissociation of LiF molecule: (a) the Mulliken charge of Li atom;
        (b) the relative total energy of LiF with respect to Li and F atoms,
        $\Delta E=E_{{\rm {LiF}}}-(E_{{\rm {Li}}}+E_{{\rm {F}}})$. LOSC2
        calculations are associated with B3LYP.}
    \label{fig:LiF} 
\end{figure}

Among the three diatomic molecules, B3LYP shows obvious delocalization
error for LiF and LiH molecules, making them good cases to demonstrate
the performance of the new SCF-LOSC approach. Thus, we highlight the
results for LiF and LiH molecules in the main text. Results for HF
molecule are documented in the supporting information. The results
for LiF are shown in Figure \ref{fig:LiF}. Because ${\rm {IP}_{{\rm {Li}}}>{\rm {EA}_{{\rm {F}}}}}$,
the LiF molecule must dissociate into neutral Li and F atom (as with
all neutral diatomic molecules). Clearly, according to Figure \ref{fig:LiF},
we see B3LYP shows significant delocalization error in electron density,
which is reflected by the positive Mulliken charge of Li atom and
underestimated dissociation energy at the dissociation limit. Based on the delocalized
electron density from B3LYP, post-LOSC-B3LYP corrects the total energy
too much and yields higher dissociation limit. In contrast to the
post-LOSC-B3LYP, the SCF-LOSC-B3LYP corrects the electron density.
As shown in Figure \ref{fig:LiF}, the Mulliken charge curve from
SCF-LOSC-B3LYP matches well with the MRCI+Q reference. In addition,
the relative total energy curve from SCF-LOSC-B3LYP almost overlaps
with the reference and shows the correct dissociation limit.

\begin{figure}[htbp]
    \centering
    \subfigure[Mulliken charge]{\includegraphics[width=0.48\linewidth]{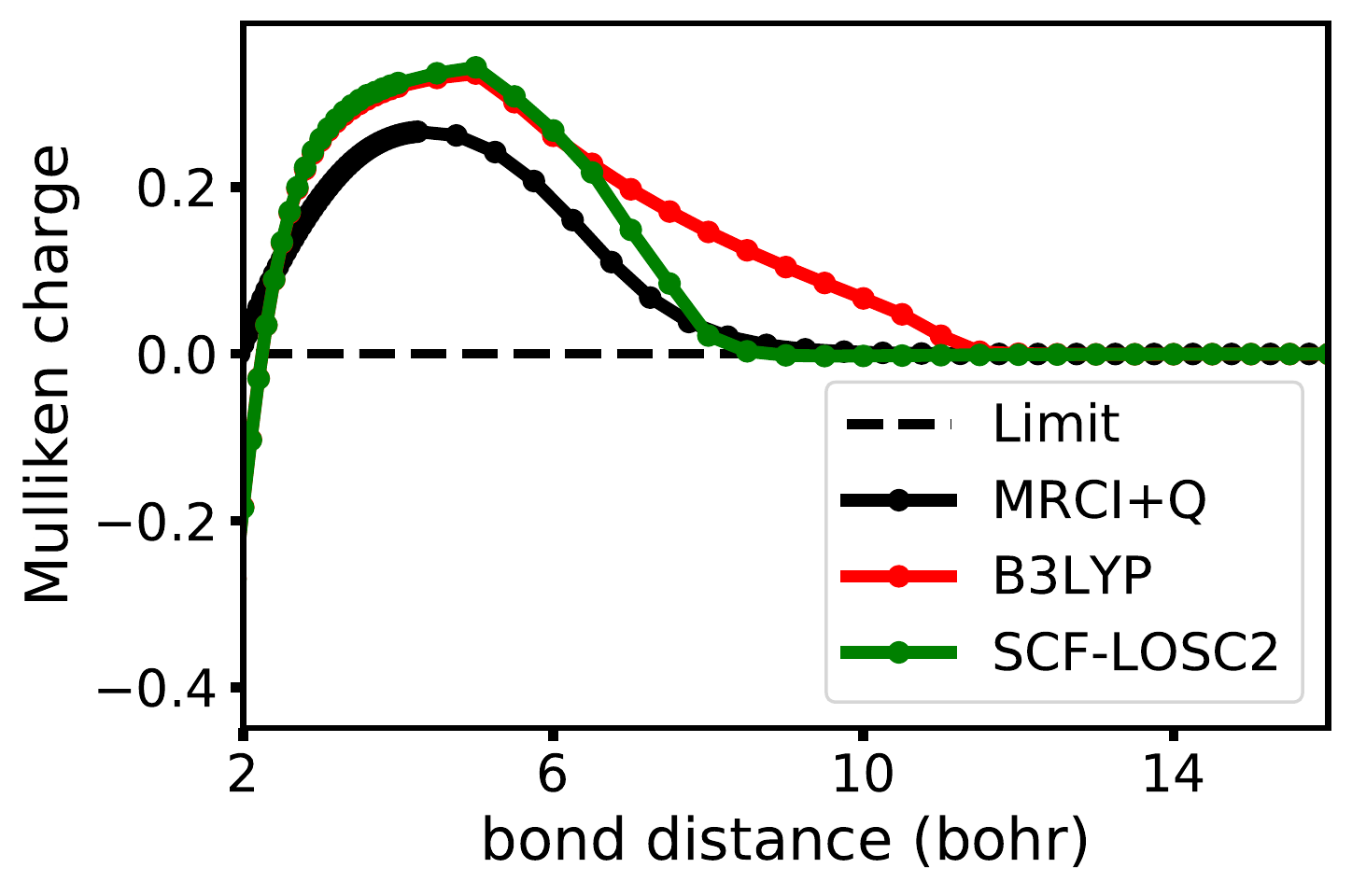}}
    \subfigure[Relative total energy]{\includegraphics[width=0.48\linewidth]{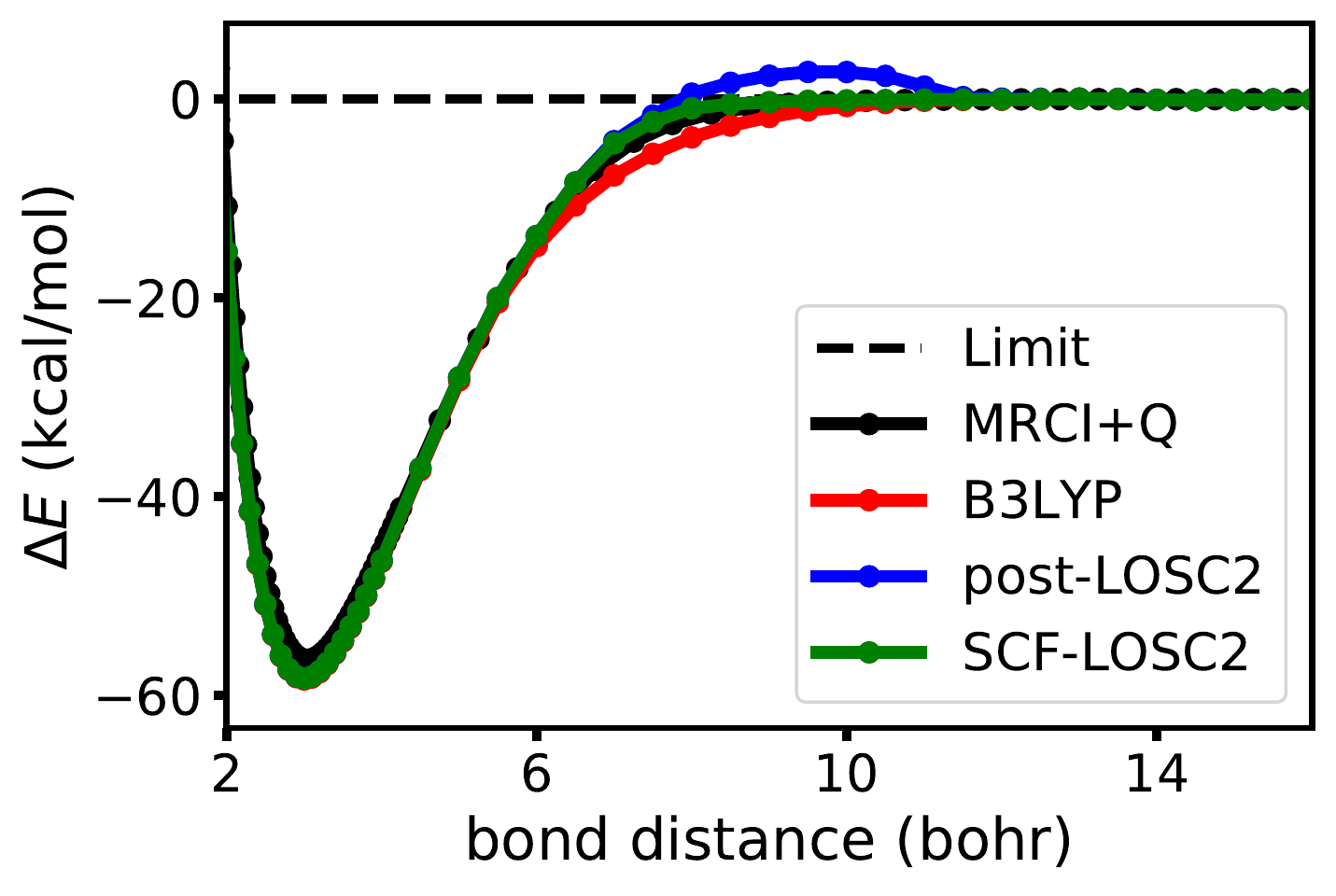}}
    \caption{Dissociation of LiH molecule: (a) the Mulliken charge of Li atom;
        (b) the relative total energy of LiH with respect to Li and H atoms,
        $\Delta E=E_{{\rm {LiH}}}-(E_{{\rm {Li}}}+E_{{\rm {H}}})$. LOSC2
        calculations are associated with B3LYP.}
    \label{fig:LiH} 
\end{figure}

Next, we look at the LiH molecule that are shown in Figure \ref{fig:LiH}.
Because ${\rm {IP}_{{\rm {Li}}}>{\rm {EA}_{{\rm {H}}}}}$, the LiH
molecule must dissociate into neutral Li and H atom. According to
Figure \ref{fig:LiH}, we notice that although B3LYP gives correct
Mulliken charge (zero charge) of Li atom at the dissociation limit,
it shows delocalization error at the bond length around 8-10 bohr,
in which the MRCI+Q gives almost zero Mulliken charge while B3LYP
gives positive charge. Such delocalized electron density (around 8-10
bohr) leads the relative total energy from B3LYP lower than the MRCI+Q
reference. In addition, the error in electron density from B3LYP
in the range of 8-10 bohr leads the post-LOSC-B3LYP to yield too high
relative total energy, showing as the small bump in the dissociation
energy curve in Figure \ref{fig:LiH}. From the results of SCF-LOSC-B3LYP,
we observe that the electron density is corrected, showing as the
Mulliken charge of Li in the range of 8-10 bohr was corrected down
to zero. With the corrected electron density from SCF-LOSC-B3LYP,
the relative total energy from SCF-LOSC-B3LYP matches much better with
the MRCI+Q reference than B3LYP and post-LOSC-B3LYP.

\begin{figure}[htbp]
    \centering \subfigure[Mulliken charge]{\includegraphics[width=0.48\linewidth]{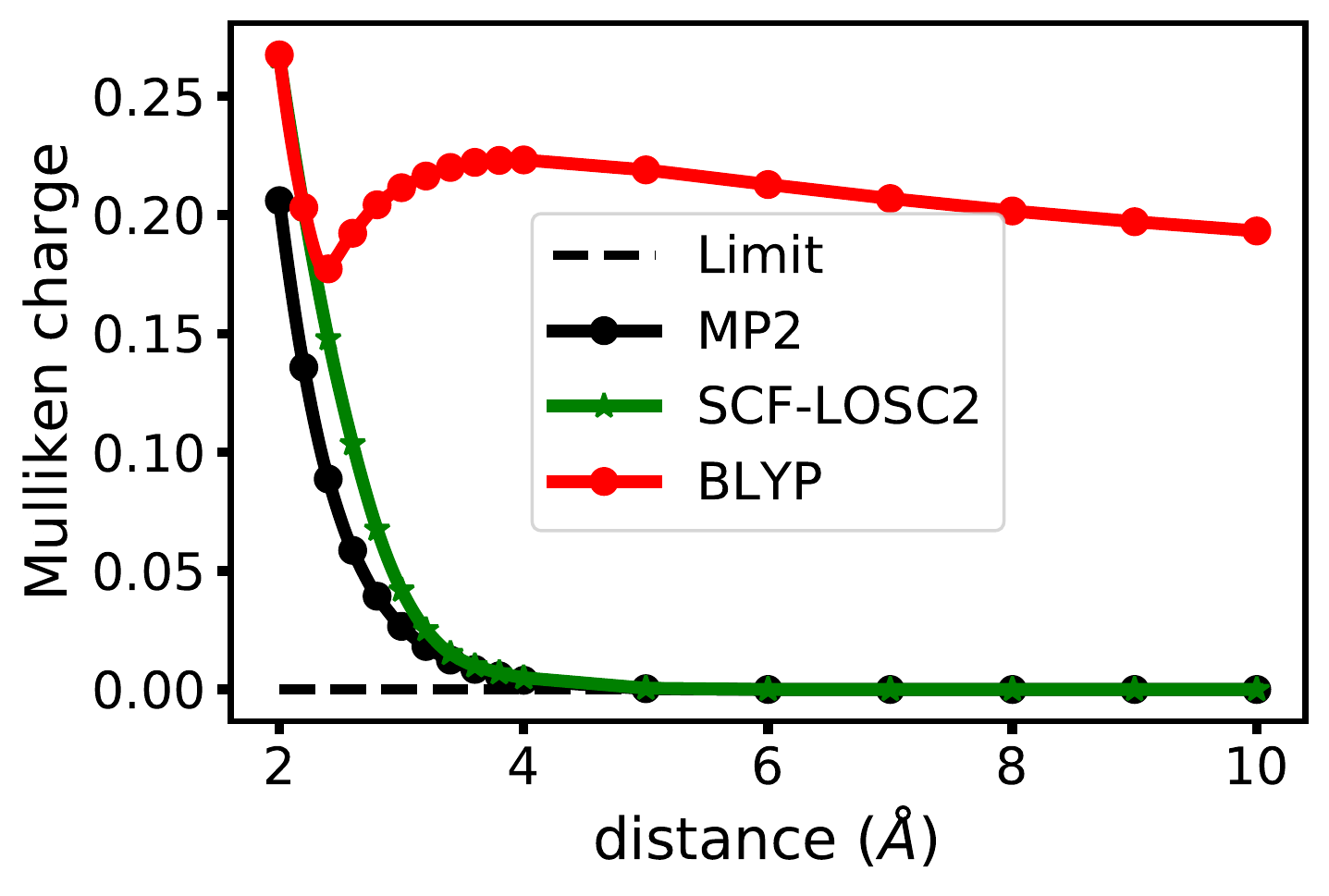}}
    \subfigure[Relative total energy]{\includegraphics[width=0.48\linewidth]{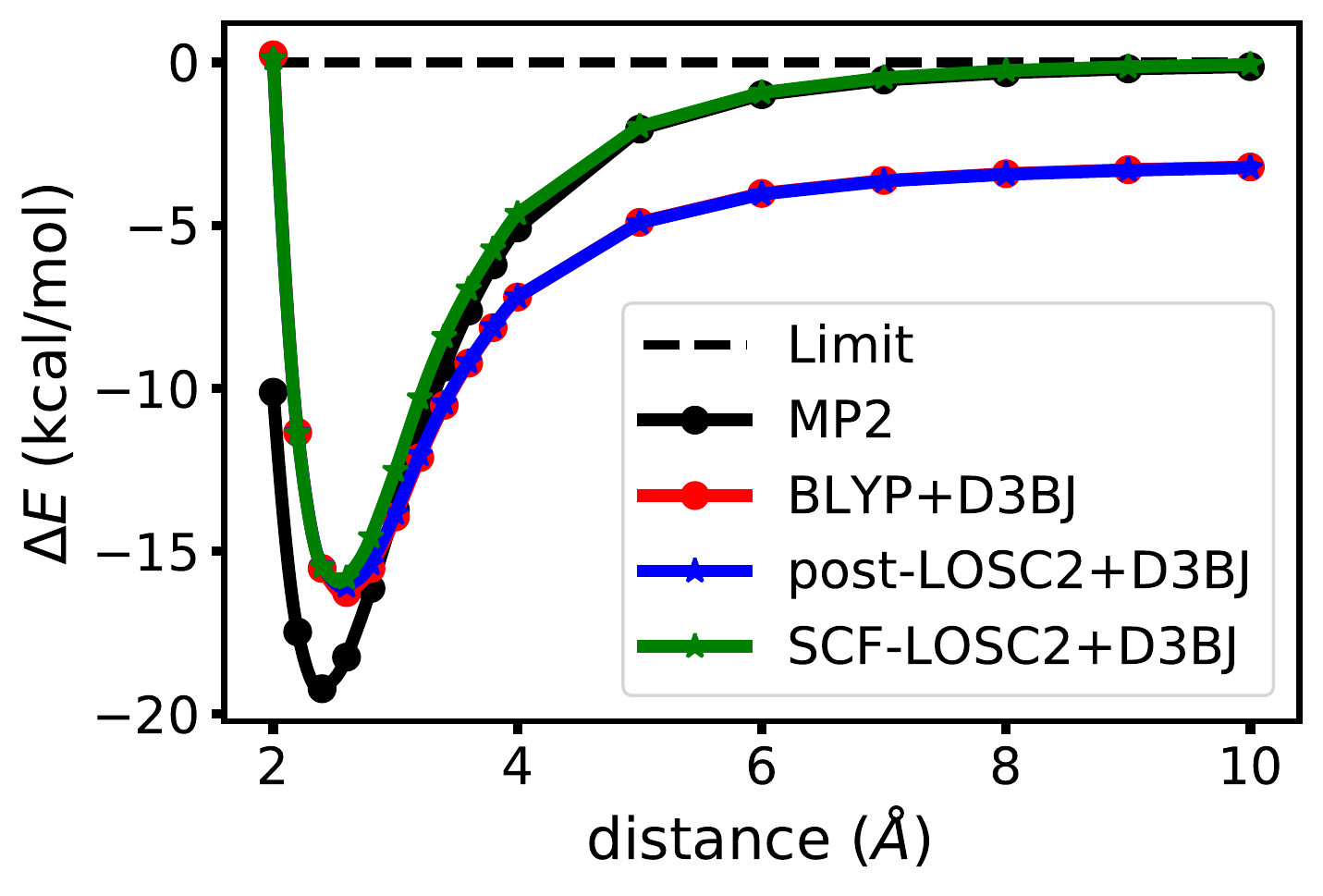}}
    \caption{Dissociation of donor-acceptor (D-A) complex (donor: 1,4-benzenediamine,
        acceptor: tetracyanoethylene (TCNE)) from different methods: (a) the
        Mulliken charge of the donor molecule; (b) the relative total energy
        of D-A complex with respect to neutral donor and acceptor molecules,
        $\Delta E=E_{{\rm {DA}}}-(E_{{\rm {D}}}+E_{{\rm {A}}})$. LOSC2 calculations
        are associated with BLYP. The D3 version of Grimme's dispersion with
        Becke-Johnson damping (D3BJ) \cite{grimmeEffectDampingFunction2011}
        from BLYP functional is added to all the DFT energy results. }
    \label{fig:DA} 
\end{figure}

We also study
a donor-acceptor (D-A) organic complex system to demonstrate the performance
of new SCF-LOSC for more complicated and larger systems. The donor
molecule is 1,4-benzenediamine and the acceptor molecule is tetracyanoethylene
(TCNE). Because ${\rm {IP_{D}}>{\rm {EA_{{\rm {A}}}}}}$, the D-A
complex will dissociate into two neutral subsystems. Figure \ref{fig:DA}
shows the Mulliken charge results and the dissociation energy from
DFT and MP2 \cite{mollerNoteApproximationTreatment1934} calculations.
BLYP is used as the parent function for DFT. Clearly, we see that
the parent functional BLYP shows delocalization error at the dissociation
limit. The donor molecule has spuriously positive charge, which means
there is partial charge transferring from the donor to the acceptor molecule
and the electron density is delocalized incorrectly. Due to the delocalized
electron density from BLYP, the dissociation energy from BLYP and
post-LOSC-BLYP shows similar error. With the correction to the electron
density, we see that applying the new SCF-LOSC gives the right Mulliken
charges and total energies along the dissociation coordinates and including
dissociation limit.

\begin{figure}[htbp]
    \subfigure[IP]{\includegraphics[width=0.48\linewidth]{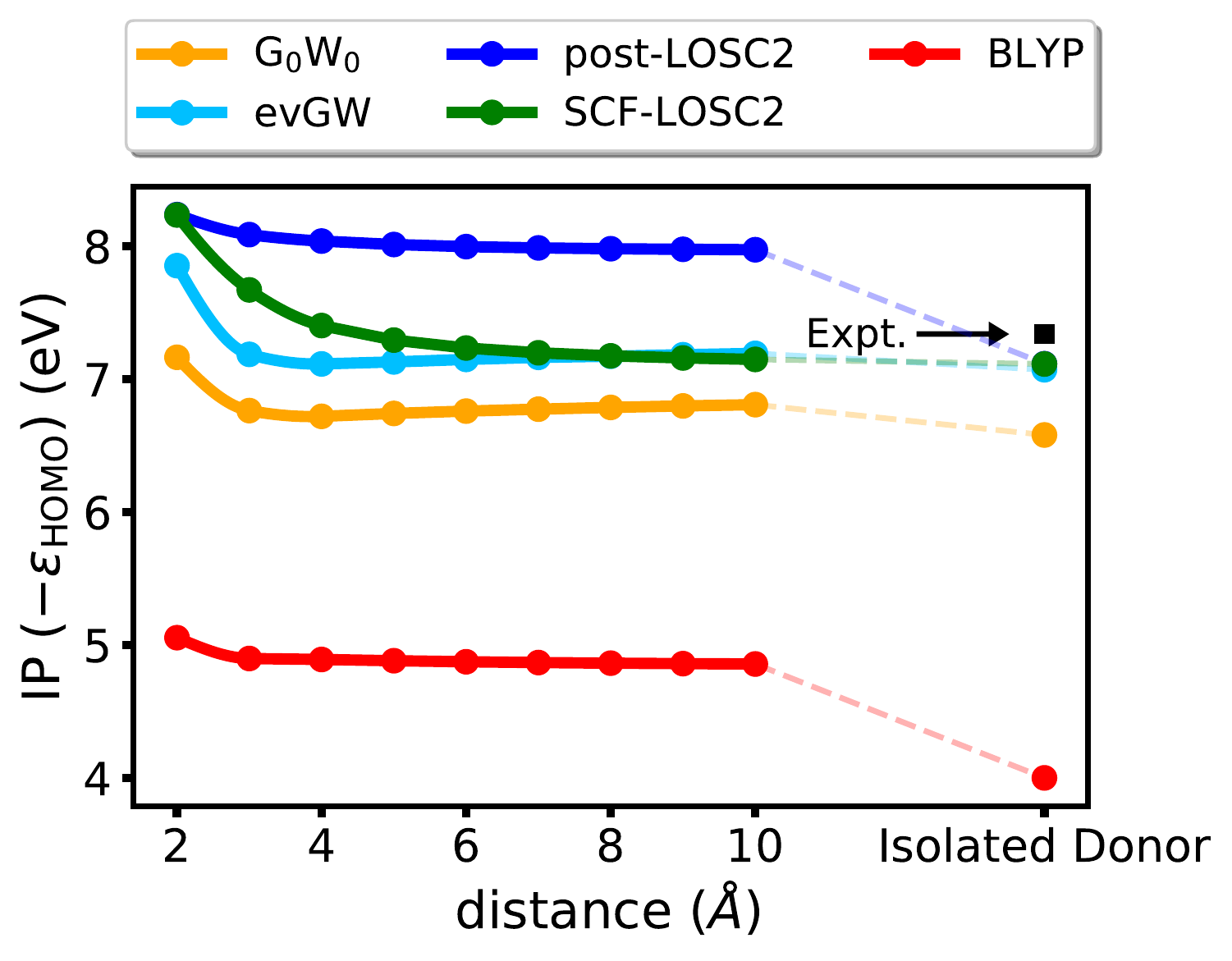}}
    \subfigure[EA]{\includegraphics[width=0.48\linewidth]{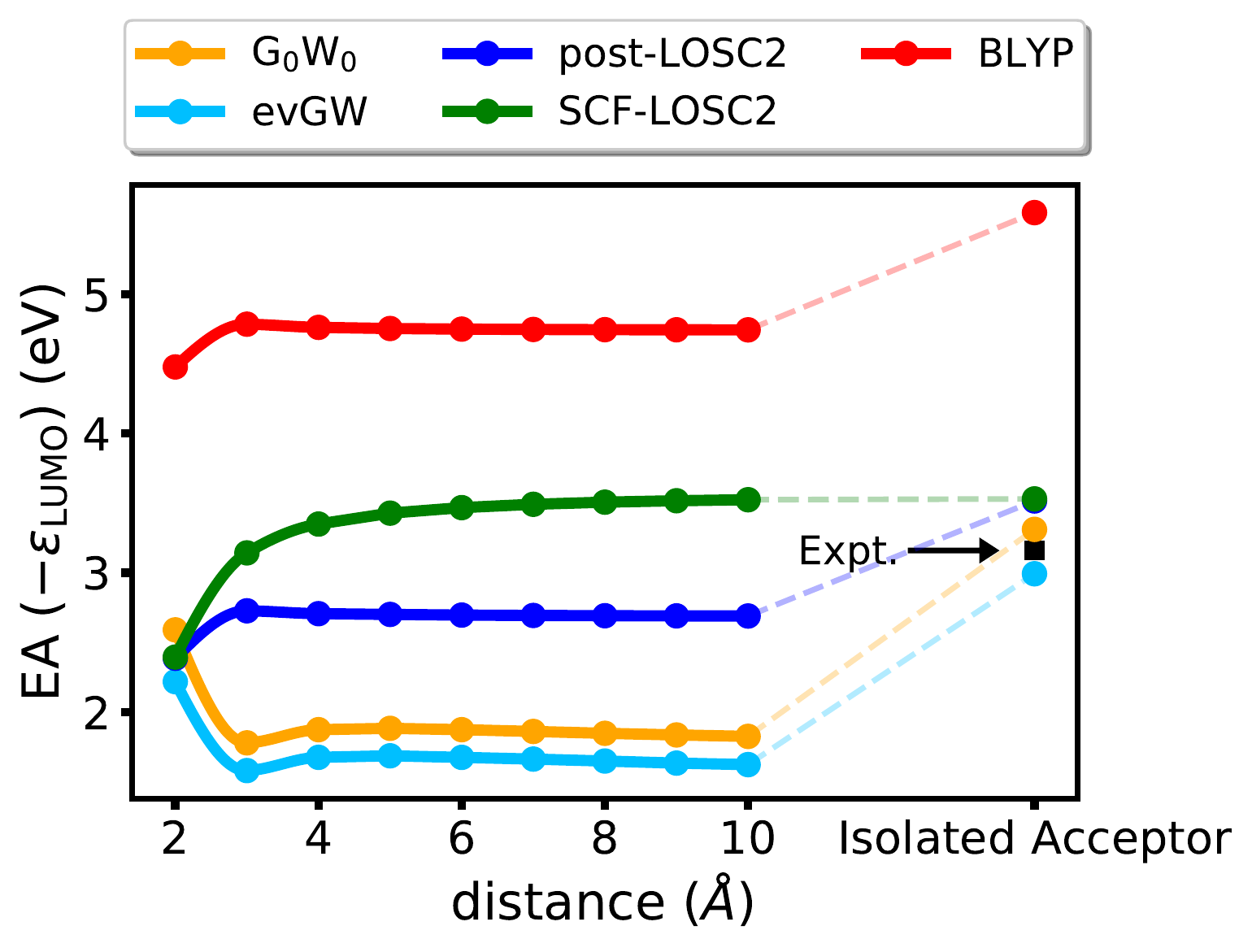}}
    \caption{HOMO and LUMO energy level alignment of donor-acceptor
        (D-A) complex (donor: 1,4-benzenediamine, acceptor: tetracyanoethylene
        (TCNE)) from different methods: (a) the first IP ($-\varepsilon_{\rm{HOMO}}$)
        of D-A complex along the separation distance from 2 to 10 $\text{\AA}$.
        At the right end of
        the figure, the first IP of the isolated donor molecule is plotted 
        to be compared with the first IP from D-A complex calculations.
        The experimental IP of the isolated donor molecule
        is 7.34 eV
        \cite{streetsMesomericMixingEnergy1972}
        and marked with an arrow.
        (b) the first EA ($-\varepsilon_{\rm{LUMO}}$)
        of D-A complex along the separation distance from 2 to 10 $\text{\AA}$.
        At the right end of the figure, the first EA of the
        isolated acceptor molecule is plotted to be compared with the first
        EA from D-A complex calculations. The experimental EA of the isolated
        acceptor molecule is 3.16 eV
        \cite{khuseynovPhotoelectronSpectroscopyPhotochemistry2012}
        and marked with an arrow.
        LOSC2 and GW calculations are based on  BLYP.}
    \label{fig:DA_2} 
\end{figure}

In addition to the Mulliken charge analysis and dissociation energy
of this D-A complex system, we  examine its energy level alignment
(the first IP and EA) along the binding distance. Figure \ref{fig:DA_2}
shows the trend of first IP and EA with respect to the change of separation
distance. The experimental IP of the donor and EA of the acceptor
molecule are plotted as the reference values for the dissociation
limit. Along the dissociation, the D-A complex system is calculated
with DFT and GW methods (G$_{0}$W$_{0}$ and eigenvalue self-consistent
GW (evGW)). For the DFT calculations, the negative orbital energy of
HOMO and the negative orbital energy of LUMO are used to evaluate
the first IP and EA respectively \cite{Cohen08}. For GW calculations,
the obtained quasiparticle energies are used to evaluate the first
IP and EA accordingly. According to Figure \ref{fig:DA_2}, we see
BLYP shows significant underestimation of IP and overestimation of
EA because of the delocalization error.
The results from GW method,
which is based on BLYP functional, are also affected by the error
in the electron density from BLYP. Especially in the case of EA, the results
from GW shows obvious underestimation. At large distances, even at
the level of evGW, the error in the EA energy level is underestimated
by about 1.5 eV, which is large. Since SCF-LOSC-BLYP corrects
the electron density, the results from SCF-LOSC-BLYP
are close to the reference value at the dissociation limit, especially
for the first IP. The results from post-LOSC-BLYP, shows much improvement
compared to the BLYP, however, they are not as good as SCF-LOSC-BLYP
results.

The description of interface charge distribution and energy
level alignment in this charge transfer system demonstrates the major
improvement from the SCF-LOSC and clearly highlights the importance
of getting correct charge distributions through self consistent calculations,
for the correct energy level alignment in DFT as well as in
Green's functional calculations.

In summary, the new SCF-LOSC calculation overcomes the convergence issue and
is very effective in practice. More importantly, we observe that the
new SCF-LOSC approach is able to produce the correct electron densities,
total energies and energy level alignment. The performance from SCF-LOSC
is more reliable than the post-LOSC, especially for the cases in which
the converged density from the parent DFA shows great delocalization
error. With the good performance and reliability, we believe the new
SCF-LOSC method would be promising to study problems related to electron
densities, and quasiparticle energy level alignment in large molecules
and interface systems.

\section{Acknowledgement}

Y. M. and Z.C. acknowledge the support from the National Institute
of General Medical Sciences of the National Institutes of Health under
award number R01-GM061870. W.Y. acknowledges the support from the
National Science Foundation (Grant No. CHE-1900338). Y. M. was also
supported by the Shaffer-Hunnicutt Fellowship and Z.C. by
the Kathleen Zielik Fellowship from Duke University.

\bibliography{FL_LOSC,my,DFT,qm4d,WY_add}

\providecommand{\latin}[1]{#1}
\providecommand*\mcitethebibliography{\thebibliography}
\csname @ifundefined\endcsname{endmcitethebibliography}
  {\let\endmcitethebibliography\endthebibliography}{}
\begin{mcitethebibliography}{8}
\providecommand*\natexlab[1]{#1}
\providecommand*\mciteSetBstSublistMode[1]{}
\providecommand*\mciteSetBstMaxWidthForm[2]{}
\providecommand*\mciteBstWouldAddEndPuncttrue
  {\def\EndOfBibitem{\unskip.}}
\providecommand*\mciteBstWouldAddEndPunctfalse
  {\let\EndOfBibitem\relax}
\providecommand*\mciteSetBstMidEndSepPunct[3]{}
\providecommand*\mciteSetBstSublistLabelBeginEnd[3]{}
\providecommand*\EndOfBibitem{}
\mciteSetBstSublistMode{f}
\mciteSetBstMaxWidthForm{subitem}{(\alph{mcitesubitemcount})}
\mciteSetBstSublistLabelBeginEnd
  {\mcitemaxwidthsubitemform\space}
  {\relax}
  {\relax}

\bibitem[qm4()]{qm4d}
An In-House Program for {{QM}}/{{MM}} Simulations. https://qm4d.org/\relax
\mciteBstWouldAddEndPuncttrue
\mciteSetBstMidEndSepPunct{\mcitedefaultmidpunct}
{\mcitedefaultendpunct}{\mcitedefaultseppunct}\relax
\EndOfBibitem
\bibitem[Gordon and Schmidt(2005)Gordon, and
  Schmidt]{gordonChapter41Advances2005}
Gordon,~M.~S.; Schmidt,~M.~W. In \emph{Theory and {{Applications}} of
  {{Computational Chemistry}}}; Dykstra,~C.~E., Frenking,~G., Kim,~K.~S.,
  Scuseria,~G.~E., Eds.; {Elsevier}: {Amsterdam}, 2005; pp 1167--1189\relax
\mciteBstWouldAddEndPuncttrue
\mciteSetBstMidEndSepPunct{\mcitedefaultmidpunct}
{\mcitedefaultendpunct}{\mcitedefaultseppunct}\relax
\EndOfBibitem
\bibitem[Schmidt \latin{et~al.}(1993)Schmidt, Baldridge, Boatz, Elbert, Gordon,
  Jensen, Koseki, Matsunaga, Nguyen, Su, Windus, Dupuis, and
  Montgomery]{schmidtGeneralAtomicMolecular1993}
Schmidt,~M.~W.; Baldridge,~K.~K.; Boatz,~J.~A.; Elbert,~S.~T.; Gordon,~M.~S.;
  Jensen,~J.~H.; Koseki,~S.; Matsunaga,~N.; Nguyen,~K.~A.; Su,~S.;
  Windus,~T.~L.; Dupuis,~M.; Montgomery,~J.~A. \emph{J. Comput. Chem.}
  \textbf{1993}, \emph{14}, 1347--1363\relax
\mciteBstWouldAddEndPuncttrue
\mciteSetBstMidEndSepPunct{\mcitedefaultmidpunct}
{\mcitedefaultendpunct}{\mcitedefaultseppunct}\relax
\EndOfBibitem
\bibitem[Blum \latin{et~al.}(2009)Blum, Gehrke, Hanke, Havu, Havu, Ren, Reuter,
  and Scheffler]{blumInitioMolecularSimulations2009}
Blum,~V.; Gehrke,~R.; Hanke,~F.; Havu,~P.; Havu,~V.; Ren,~X.; Reuter,~K.;
  Scheffler,~M. \emph{Comput. Phys. Commun.} \textbf{2009}, \emph{180},
  2175--2196\relax
\mciteBstWouldAddEndPuncttrue
\mciteSetBstMidEndSepPunct{\mcitedefaultmidpunct}
{\mcitedefaultendpunct}{\mcitedefaultseppunct}\relax
\EndOfBibitem
\bibitem[Ren \latin{et~al.}(2012)Ren, Rinke, Blum, Wieferink, Tkatchenko,
  Sanfilippo, Reuter, and
  Scheffler]{renResolutionofidentityApproachHartree2012}
Ren,~X.; Rinke,~P.; Blum,~V.; Wieferink,~J.; Tkatchenko,~A.; Sanfilippo,~A.;
  Reuter,~K.; Scheffler,~M. \emph{New J. Phys.} \textbf{2012}, \emph{14},
  053020\relax
\mciteBstWouldAddEndPuncttrue
\mciteSetBstMidEndSepPunct{\mcitedefaultmidpunct}
{\mcitedefaultendpunct}{\mcitedefaultseppunct}\relax
\EndOfBibitem
\bibitem[Frisch \latin{et~al.}(2016)Frisch, Trucks, Schlegel, Scuseria, Robb,
  Cheeseman, Scalmani, Barone, Petersson, Nakatsuji, Li, Caricato, Marenich,
  Bloino, Janesko, Gomperts, Mennucci, Hratchian, Ortiz, Izmaylov, Sonnenberg,
  Williams-Young, Ding, Lipparini, Egidi, Goings, Peng, Petrone, Henderson,
  Ranasinghe, Zakrzewski, Gao, Rega, Zheng, Liang, Hada, Ehara, Toyota, Fukuda,
  Hasegawa, Ishida, Nakajima, Honda, Kitao, Nakai, Vreven, Throssell,
  Montgomery, Peralta, Ogliaro, Bearpark, Heyd, Brothers, Kudin, Staroverov,
  Keith, Kobayashi, Normand, Raghavachari, Rendell, Burant, Iyengar, Tomasi,
  Cossi, Millam, Klene, Adamo, Cammi, Ochterski, Martin, Morokuma, Farkas,
  Foresman, and Fox]{g16}
Frisch,~M.~J.; Trucks,~G.~W.; Schlegel,~H.~B.; Scuseria,~G.~E.; Robb,~M.~A.;
  Cheeseman,~J.~R.; Scalmani,~G.; Barone,~V.; Petersson,~G.~A.; Nakatsuji,~H.;
  Li,~X.; Caricato,~M.; Marenich,~A.~V.; Bloino,~J.; Janesko,~B.~G.;
  Gomperts,~R.; Mennucci,~B.; Hratchian,~H.~P.; Ortiz,~J.~V.; Izmaylov,~A.~F.;
  Sonnenberg,~J.~L.; Williams-Young,~D.; Ding,~F.; Lipparini,~F.; Egidi,~F.;
  Goings,~J.; Peng,~B.; Petrone,~A.; Henderson,~T.; Ranasinghe,~D.;
  Zakrzewski,~V.~G.; Gao,~J.; Rega,~N.; Zheng,~G.; Liang,~W.; Hada,~M.;
  Ehara,~M.; Toyota,~K.; Fukuda,~R.; Hasegawa,~J.; Ishida,~M.; Nakajima,~T.;
  Honda,~Y.; Kitao,~O.; Nakai,~H.; Vreven,~T.; Throssell,~K.;
  Montgomery,~J.~A.,~{Jr.}; Peralta,~J.~E.; Ogliaro,~F.; Bearpark,~M.~J.;
  Heyd,~J.~J.; Brothers,~E.~N.; Kudin,~K.~N.; Staroverov,~V.~N.; Keith,~T.~A.;
  Kobayashi,~R.; Normand,~J.; Raghavachari,~K.; Rendell,~A.~P.; Burant,~J.~C.;
  Iyengar,~S.~S.; Tomasi,~J.; Cossi,~M.; Millam,~J.~M.; Klene,~M.; Adamo,~C.;
  Cammi,~R.; Ochterski,~J.~W.; Martin,~R.~L.; Morokuma,~K.; Farkas,~O.;
  Foresman,~J.~B.; Fox,~D.~J. Gaussian 16 {R}evision {A}.03. 2016; Gaussian
  Inc. Wallingford CT\relax
\mciteBstWouldAddEndPuncttrue
\mciteSetBstMidEndSepPunct{\mcitedefaultmidpunct}
{\mcitedefaultendpunct}{\mcitedefaultseppunct}\relax
\EndOfBibitem
\bibitem[Su \latin{et~al.}(2020)Su, Mahler, and
  Yang]{suPreservingSymmetryDegeneracy2020a}
Su,~N.~Q.; Mahler,~A.; Yang,~W. \emph{J. Phys. Chem. Lett.} \textbf{2020},
  \emph{11}, 1528--1535\relax
\mciteBstWouldAddEndPuncttrue
\mciteSetBstMidEndSepPunct{\mcitedefaultmidpunct}
{\mcitedefaultendpunct}{\mcitedefaultseppunct}\relax
\EndOfBibitem
\end{mcitethebibliography}


\providecommand{\latin}[1]{#1}
\providecommand*\mcitethebibliography{\thebibliography}
\csname @ifundefined\endcsname{endmcitethebibliography}
  {\let\endmcitethebibliography\endthebibliography}{}
\begin{mcitethebibliography}{62}
\providecommand*\natexlab[1]{#1}
\providecommand*\mciteSetBstSublistMode[1]{}
\providecommand*\mciteSetBstMaxWidthForm[2]{}
\providecommand*\mciteBstWouldAddEndPuncttrue
  {\def\EndOfBibitem{\unskip.}}
\providecommand*\mciteBstWouldAddEndPunctfalse
  {\let\EndOfBibitem\relax}
\providecommand*\mciteSetBstMidEndSepPunct[3]{}
\providecommand*\mciteSetBstSublistLabelBeginEnd[3]{}
\providecommand*\EndOfBibitem{}
\mciteSetBstSublistMode{f}
\mciteSetBstMaxWidthForm{subitem}{(\alph{mcitesubitemcount})}
\mciteSetBstSublistLabelBeginEnd
  {\mcitemaxwidthsubitemform\space}
  {\relax}
  {\relax}

\bibitem[Hohenberg and Kohn(1964)Hohenberg, and
  Kohn]{hohenbergInhomogeneousElectronGas1964b}
Hohenberg,~P.; Kohn,~W. Inhomogeneous {{Electron Gas}}. \emph{Phys. Rev.}
  \textbf{1964}, \emph{136}, B864--B871\relax
\mciteBstWouldAddEndPuncttrue
\mciteSetBstMidEndSepPunct{\mcitedefaultmidpunct}
{\mcitedefaultendpunct}{\mcitedefaultseppunct}\relax
\EndOfBibitem
\bibitem[Kohn and Sham(1965)Kohn, and
  Sham]{kohnSelfConsistentEquationsIncluding1965b}
Kohn,~W.; Sham,~L.~J. Self-{{Consistent Equations Including Exchange}} and
  {{Correlation Effects}}. \emph{Phys. Rev.} \textbf{1965}, \emph{140},
  A1133--A1138\relax
\mciteBstWouldAddEndPuncttrue
\mciteSetBstMidEndSepPunct{\mcitedefaultmidpunct}
{\mcitedefaultendpunct}{\mcitedefaultseppunct}\relax
\EndOfBibitem
\bibitem[Parr and Yang(1994)Parr, and
  Yang]{parrDensityFunctionalTheoryAtoms1994}
Parr,~R.~G.; Yang,~W. \emph{Density-{{Functional Theory}} of {{Atoms}} and
  {{Molecules}}}; International {{Series}} of {{Monographs}} on {{Chemistry}};
  {Oxford University Press}: {Oxford, New York}, 1994\relax
\mciteBstWouldAddEndPuncttrue
\mciteSetBstMidEndSepPunct{\mcitedefaultmidpunct}
{\mcitedefaultendpunct}{\mcitedefaultseppunct}\relax
\EndOfBibitem
\bibitem[Vosko \latin{et~al.}(1980)Vosko, Wilk, and
  Nusair]{voskoAccurateSpindependentElectron1980a}
Vosko,~S.~H.; Wilk,~L.; Nusair,~M. Accurate Spin-Dependent Electron Liquid
  Correlation Energies for Local Spin Density Calculations: A Critical
  Analysis. \emph{Can. J. Phys.} \textbf{1980}, \emph{58}, 1200--1211\relax
\mciteBstWouldAddEndPuncttrue
\mciteSetBstMidEndSepPunct{\mcitedefaultmidpunct}
{\mcitedefaultendpunct}{\mcitedefaultseppunct}\relax
\EndOfBibitem
\bibitem[Perdew and Wang(1992)Perdew, and
  Wang]{perdewAccurateSimpleAnalytic1992}
Perdew,~J.~P.; Wang,~Y. Accurate and Simple Analytic Representation of the
  Electron-Gas Correlation Energy. \emph{Phys. Rev. B} \textbf{1992},
  \emph{45}, 13244--13249\relax
\mciteBstWouldAddEndPuncttrue
\mciteSetBstMidEndSepPunct{\mcitedefaultmidpunct}
{\mcitedefaultendpunct}{\mcitedefaultseppunct}\relax
\EndOfBibitem
\bibitem[Becke(1988)]{beckeDensityfunctionalExchangeenergyApproximation1988a}
Becke,~A.~D. Density-Functional Exchange-Energy Approximation with Correct
  Asymptotic Behavior. \emph{Phys. Rev. A} \textbf{1988}, \emph{38},
  3098--3100\relax
\mciteBstWouldAddEndPuncttrue
\mciteSetBstMidEndSepPunct{\mcitedefaultmidpunct}
{\mcitedefaultendpunct}{\mcitedefaultseppunct}\relax
\EndOfBibitem
\bibitem[Lee \latin{et~al.}(1988)Lee, Yang, and
  Parr]{leeDevelopmentColleSalvettiCorrelationenergy1988a}
Lee,~C.; Yang,~W.; Parr,~R.~G. Development of the {{Colle}}-{{Salvetti}}
  Correlation-Energy Formula into a Functional of the Electron Density.
  \emph{Phys. Rev. B} \textbf{1988}, \emph{37}, 785--789\relax
\mciteBstWouldAddEndPuncttrue
\mciteSetBstMidEndSepPunct{\mcitedefaultmidpunct}
{\mcitedefaultendpunct}{\mcitedefaultseppunct}\relax
\EndOfBibitem
\bibitem[Perdew \latin{et~al.}(1996)Perdew, Burke, and
  Ernzerhof]{perdewGeneralizedGradientApproximation1996}
Perdew,~J.~P.; Burke,~K.; Ernzerhof,~M. Generalized {{Gradient Approximation
  Made Simple}}. \emph{Phys. Rev. Lett.} \textbf{1996}, \emph{77},
  3865--3868\relax
\mciteBstWouldAddEndPuncttrue
\mciteSetBstMidEndSepPunct{\mcitedefaultmidpunct}
{\mcitedefaultendpunct}{\mcitedefaultseppunct}\relax
\EndOfBibitem
\bibitem[Stephens \latin{et~al.}(1994)Stephens, Devlin, Chabalowski, and
  Frisch]{stephensInitioCalculationVibrational1994}
Stephens,~P.~J.; Devlin,~F.~J.; Chabalowski,~C.~F.; Frisch,~M.~J. Ab {{Initio
  Calculation}} of {{Vibrational Absorption}} and {{Circular Dichroism Spectra
  Using Density Functional Force Fields}}. \emph{J. Phys. Chem.} \textbf{1994},
  \emph{98}, 11623--11627\relax
\mciteBstWouldAddEndPuncttrue
\mciteSetBstMidEndSepPunct{\mcitedefaultmidpunct}
{\mcitedefaultendpunct}{\mcitedefaultseppunct}\relax
\EndOfBibitem
\bibitem[Adamo and Barone(1999)Adamo, and
  Barone]{adamoReliableDensityFunctional1999}
Adamo,~C.; Barone,~V. Toward Reliable Density Functional Methods without
  Adjustable Parameters: {{The PBE0}} Model. \emph{J. Chem. Phys.}
  \textbf{1999}, \emph{110}, 6158--6170\relax
\mciteBstWouldAddEndPuncttrue
\mciteSetBstMidEndSepPunct{\mcitedefaultmidpunct}
{\mcitedefaultendpunct}{\mcitedefaultseppunct}\relax
\EndOfBibitem
\bibitem[Ernzerhof and Scuseria(1999)Ernzerhof, and
  Scuseria]{ernzerhofAssessmentPerdewBurke1999}
Ernzerhof,~M.; Scuseria,~G.~E. Assessment of the
  {{Perdew}}\textendash{{Burke}}\textendash{{Ernzerhof}} Exchange-Correlation
  Functional. \emph{J. Chem. Phys.} \textbf{1999}, \emph{110}, 5029--5036\relax
\mciteBstWouldAddEndPuncttrue
\mciteSetBstMidEndSepPunct{\mcitedefaultmidpunct}
{\mcitedefaultendpunct}{\mcitedefaultseppunct}\relax
\EndOfBibitem
\bibitem[{Mori-S{\'a}nchez} \latin{et~al.}(2008){Mori-S{\'a}nchez}, Cohen, and
  Yang]{mori-sanchezLocalizationDelocalizationErrors2008c}
{Mori-S{\'a}nchez},~P.; Cohen,~A.~J.; Yang,~W. Localization and
  {{Delocalization Errors}} in {{Density Functional Theory}} and
  {{Implications}} for {{Band}}-{{Gap Prediction}}. \emph{Phys. Rev. Lett.}
  \textbf{2008}, \emph{100}, 146401\relax
\mciteBstWouldAddEndPuncttrue
\mciteSetBstMidEndSepPunct{\mcitedefaultmidpunct}
{\mcitedefaultendpunct}{\mcitedefaultseppunct}\relax
\EndOfBibitem
\bibitem[Cohen \latin{et~al.}(2008)Cohen, {Mori-S{\'a}nchez}, and
  Yang]{cohenFractionalChargePerspective2008b}
Cohen,~A.~J.; {Mori-S{\'a}nchez},~P.; Yang,~W. Fractional Charge Perspective on
  the Band Gap in Density-Functional Theory. \emph{Phys. Rev. B} \textbf{2008},
  \emph{77}, 115123\relax
\mciteBstWouldAddEndPuncttrue
\mciteSetBstMidEndSepPunct{\mcitedefaultmidpunct}
{\mcitedefaultendpunct}{\mcitedefaultseppunct}\relax
\EndOfBibitem
\bibitem[Cohen \latin{et~al.}(2008)Cohen, {Mori-S{\'a}nchez}, and
  Yang]{cohenInsightsCurrentLimitations2008b}
Cohen,~A.~J.; {Mori-S{\'a}nchez},~P.; Yang,~W. Insights into {{Current
  Limitations}} of {{Density Functional Theory}}. \emph{Science} \textbf{2008},
  \emph{321}, 792--794\relax
\mciteBstWouldAddEndPuncttrue
\mciteSetBstMidEndSepPunct{\mcitedefaultmidpunct}
{\mcitedefaultendpunct}{\mcitedefaultseppunct}\relax
\EndOfBibitem
\bibitem[Cohen \latin{et~al.}(2012)Cohen, {Mori-S{\'a}nchez}, and
  Yang]{cohenChallengesDensityFunctional2012a}
Cohen,~A.~J.; {Mori-S{\'a}nchez},~P.; Yang,~W. Challenges for {{Density
  Functional Theory}}. \emph{Chem. Rev.} \textbf{2012}, \emph{112},
  289--320\relax
\mciteBstWouldAddEndPuncttrue
\mciteSetBstMidEndSepPunct{\mcitedefaultmidpunct}
{\mcitedefaultendpunct}{\mcitedefaultseppunct}\relax
\EndOfBibitem
\bibitem[Li \latin{et~al.}(2018)Li, Zheng, Su, and
  Yang]{liLocalizedOrbitalScaling2018c}
Li,~C.; Zheng,~X.; Su,~N.~Q.; Yang,~W. Localized Orbital Scaling Correction for
  Systematic Elimination of Delocalization Error in Density Functional
  Approximations. \emph{Natl. Sci. Rev.} \textbf{2018}, \emph{5},
  203--215\relax
\mciteBstWouldAddEndPuncttrue
\mciteSetBstMidEndSepPunct{\mcitedefaultmidpunct}
{\mcitedefaultendpunct}{\mcitedefaultseppunct}\relax
\EndOfBibitem
\bibitem[Perdew \latin{et~al.}(1982)Perdew, Parr, Levy, and
  Balduz]{perdewDensityFunctionalTheoryFractional1982a}
Perdew,~J.~P.; Parr,~R.~G.; Levy,~M.; Balduz,~J.~L. Density-{{Functional
  Theory}} for {{Fractional Particle Number}}: {{Derivative Discontinuities}}
  of the {{Energy}}. \emph{Phys. Rev. Lett.} \textbf{1982}, \emph{49},
  1691--1694\relax
\mciteBstWouldAddEndPuncttrue
\mciteSetBstMidEndSepPunct{\mcitedefaultmidpunct}
{\mcitedefaultendpunct}{\mcitedefaultseppunct}\relax
\EndOfBibitem
\bibitem[Yang \latin{et~al.}(2000)Yang, Zhang, and
  Ayers]{yangDegenerateGroundStates2000b}
Yang,~W.; Zhang,~Y.; Ayers,~P.~W. Degenerate {{Ground States}} and a
  {{Fractional Number}} of {{Electrons}} in {{Density}} and {{Reduced Density
  Matrix Functional Theory}}. \emph{Phys. Rev. Lett.} \textbf{2000}, \emph{84},
  5172--5175\relax
\mciteBstWouldAddEndPuncttrue
\mciteSetBstMidEndSepPunct{\mcitedefaultmidpunct}
{\mcitedefaultendpunct}{\mcitedefaultseppunct}\relax
\EndOfBibitem
\bibitem[Zhang and Yang(2001)Zhang, and
  Yang]{zhangPerspectiveDensityfunctionalTheory2001}
Zhang,~Y.; Yang,~W. In \emph{Theoretical {{Chemistry Accounts}}: {{New Century
  Issue}}}; Cramer,~C.~J., Truhlar,~D.~G., Eds.; {Springer}: {Berlin,
  Heidelberg}, 2001; pp 346--348\relax
\mciteBstWouldAddEndPuncttrue
\mciteSetBstMidEndSepPunct{\mcitedefaultmidpunct}
{\mcitedefaultendpunct}{\mcitedefaultseppunct}\relax
\EndOfBibitem
\bibitem[Yang \latin{et~al.}(2012)Yang, Cohen, and
  {Mori-S{\'a}nchez}]{yangDerivativeDiscontinuityBandgap2012a}
Yang,~W.; Cohen,~A.~J.; {Mori-S{\'a}nchez},~P. Derivative Discontinuity,
  Bandgap and Lowest Unoccupied Molecular Orbital in Density Functional Theory.
  \emph{J. Chem. Phys.} \textbf{2012}, \emph{136}, 204111\relax
\mciteBstWouldAddEndPuncttrue
\mciteSetBstMidEndSepPunct{\mcitedefaultmidpunct}
{\mcitedefaultendpunct}{\mcitedefaultseppunct}\relax
\EndOfBibitem
\bibitem[Dutoi and {Head-Gordon}(2006)Dutoi, and
  {Head-Gordon}]{dutoiSelfinteractionErrorLocal2006}
Dutoi,~A.~D.; {Head-Gordon},~M. Self-Interaction Error of Local Density
  Functionals for Alkali\textendash Halide Dissociation. \emph{Chem. Phys.
  Lett.} \textbf{2006}, \emph{422}, 230--233\relax
\mciteBstWouldAddEndPuncttrue
\mciteSetBstMidEndSepPunct{\mcitedefaultmidpunct}
{\mcitedefaultendpunct}{\mcitedefaultseppunct}\relax
\EndOfBibitem
\bibitem[{Mori-S{\'a}nchez} \latin{et~al.}(2006){Mori-S{\'a}nchez}, Cohen, and
  Yang]{mori-sanchezManyelectronSelfinteractionError2006}
{Mori-S{\'a}nchez},~P.; Cohen,~A.~J.; Yang,~W. Many-Electron Self-Interaction
  Error in Approximate Density Functionals. \emph{J. Chem. Phys.}
  \textbf{2006}, \emph{125}, 201102\relax
\mciteBstWouldAddEndPuncttrue
\mciteSetBstMidEndSepPunct{\mcitedefaultmidpunct}
{\mcitedefaultendpunct}{\mcitedefaultseppunct}\relax
\EndOfBibitem
\bibitem[Ruzsinszky \latin{et~al.}(2006)Ruzsinszky, Perdew, Csonka, Vydrov, and
  Scuseria]{ruzsinszkySpuriousFractionalCharge2006}
Ruzsinszky,~A.; Perdew,~J.~P.; Csonka,~G.~I.; Vydrov,~O.~A.; Scuseria,~G.~E.
  Spurious Fractional Charge on Dissociated Atoms: {{Pervasive}} and Resilient
  Self-Interaction Error of Common Density Functionals. \emph{J. Chem. Phys.}
  \textbf{2006}, \emph{125}, 194112\relax
\mciteBstWouldAddEndPuncttrue
\mciteSetBstMidEndSepPunct{\mcitedefaultmidpunct}
{\mcitedefaultendpunct}{\mcitedefaultseppunct}\relax
\EndOfBibitem
\bibitem[Vydrov \latin{et~al.}(2007)Vydrov, Scuseria, and
  Perdew]{vydrovTestsFunctionalsSystems2007}
Vydrov,~O.~A.; Scuseria,~G.~E.; Perdew,~J.~P. Tests of Functionals for Systems
  with Fractional Electron Number. \emph{J. Chem. Phys.} \textbf{2007},
  \emph{126}, 154109\relax
\mciteBstWouldAddEndPuncttrue
\mciteSetBstMidEndSepPunct{\mcitedefaultmidpunct}
{\mcitedefaultendpunct}{\mcitedefaultseppunct}\relax
\EndOfBibitem
\bibitem[Zheng \latin{et~al.}(2012)Zheng, Liu, Johnson, {Contreras-Garc{\'i}a},
  and Yang]{zhengDelocalizationErrorDensityfunctional2012}
Zheng,~X.; Liu,~M.; Johnson,~E.~R.; {Contreras-Garc{\'i}a},~J.; Yang,~W.
  Delocalization Error of Density-Functional Approximations: {{A}} Distinct
  Manifestation in Hydrogen Molecular Chains. \emph{J. Chem. Phys.}
  \textbf{2012}, \emph{137}, 214106\relax
\mciteBstWouldAddEndPuncttrue
\mciteSetBstMidEndSepPunct{\mcitedefaultmidpunct}
{\mcitedefaultendpunct}{\mcitedefaultseppunct}\relax
\EndOfBibitem
\bibitem[Flores \latin{et~al.}(2009)Flores, Ortega, and
  V{\'a}zquez]{Flores098658}
Flores,~F.; Ortega,~J.; V{\'a}zquez,~H. Modelling Energy Level Alignment at
  Organic Interfaces and Density Functional Theory. \emph{Phys. Chem. Chem.
  Phys.} \textbf{2009}, \emph{11}, 8658--8675\relax
\mciteBstWouldAddEndPuncttrue
\mciteSetBstMidEndSepPunct{\mcitedefaultmidpunct}
{\mcitedefaultendpunct}{\mcitedefaultseppunct}\relax
\EndOfBibitem
\bibitem[Souza \latin{et~al.}(2013)Souza, Rungger, Pemmaraju,
  Schwingenschloegl, and Sanvito]{Souza13165112}
Souza,~A.~M.; Rungger,~I.; Pemmaraju,~C.~D.; Schwingenschloegl,~U.; Sanvito,~S.
  Constrained-{DFT} method for accurate energy-level alignment of
  metal/molecule interfaces. \emph{Phys. Rev. B} \textbf{2013}, \emph{88},
  165112\relax
\mciteBstWouldAddEndPuncttrue
\mciteSetBstMidEndSepPunct{\mcitedefaultmidpunct}
{\mcitedefaultendpunct}{\mcitedefaultseppunct}\relax
\EndOfBibitem
\bibitem[Pacchioni(2015)]{Pacchioni1580}
Pacchioni,~G. First {Principles} {Calculations} on {Oxide}-{Based}
  {Heterogeneous} {Catalysts} and {Photocatalysts}: {Problems} and {Advances}.
  \emph{Catal. Lett.} \textbf{2015}, \emph{145}, 80--94\relax
\mciteBstWouldAddEndPuncttrue
\mciteSetBstMidEndSepPunct{\mcitedefaultmidpunct}
{\mcitedefaultendpunct}{\mcitedefaultseppunct}\relax
\EndOfBibitem
\bibitem[Egger \latin{et~al.}(2015)Egger, Liu, Neaton, and Kronik]{Egger152448}
Egger,~D.~A.; Liu,~Z.-F.; Neaton,~J.~B.; Kronik,~L. Reliable {Energy} {Level}
  {Alignment} at {Physisorbed} {Molecule}{\textendash}{Metal} {Interfaces} from
  {Density} {Functional} {Theory}. \emph{Nano Lett.} \textbf{2015}, \emph{15},
  2448--2455\relax
\mciteBstWouldAddEndPuncttrue
\mciteSetBstMidEndSepPunct{\mcitedefaultmidpunct}
{\mcitedefaultendpunct}{\mcitedefaultseppunct}\relax
\EndOfBibitem
\bibitem[Wang \latin{et~al.}(2018)Wang, Sakurai, Wen, and Qi]{Wang181800260}
Wang,~S.; Sakurai,~T.; Wen,~W.; Qi,~Y. Energy {Level} {Alignment} at
  {Interfaces} in {Metal} {Halide} {Perovskite} {Solar} {Cells}. \emph{Adv.
  Mater. Interfaces} \textbf{2018}, \emph{5}, 1800260\relax
\mciteBstWouldAddEndPuncttrue
\mciteSetBstMidEndSepPunct{\mcitedefaultmidpunct}
{\mcitedefaultendpunct}{\mcitedefaultseppunct}\relax
\EndOfBibitem
\bibitem[Hegner \latin{et~al.}(2017)Hegner, Cardenas-Morcoso, Gim{\'e}nez,
  L{\'o}pez, and Galan-Mascaros]{hegnerLevelAlignmentDescriptor2017}
Hegner,~F.~S.; Cardenas-Morcoso,~D.; Gim{\'e}nez,~S.; L{\'o}pez,~N.;
  Galan-Mascaros,~J.~R. Level {Alignment} as {Descriptor} for
  {Semiconductor}/{Catalyst} {Systems} in {Water} {Splitting}: {The} {Case} of
  {Hematite}/{Cobalt} {Hexacyanoferrate} {Photoanodes}. \emph{ChemSusChem}
  \textbf{2017}, \emph{10}, 4552--4560\relax
\mciteBstWouldAddEndPuncttrue
\mciteSetBstMidEndSepPunct{\mcitedefaultmidpunct}
{\mcitedefaultendpunct}{\mcitedefaultseppunct}\relax
\EndOfBibitem
\bibitem[Savin and Flad(1995)Savin, and Flad]{Savin95327}
Savin,~A.; Flad,~H. Density {Functionals} for the {Yukawa}
  {Electron}-{Electron} {Interaction}. \emph{Int. J. Quantum Chem.}
  \textbf{1995}, \emph{56}, 327--332\relax
\mciteBstWouldAddEndPuncttrue
\mciteSetBstMidEndSepPunct{\mcitedefaultmidpunct}
{\mcitedefaultendpunct}{\mcitedefaultseppunct}\relax
\EndOfBibitem
\bibitem[Gill \latin{et~al.}(1996)Gill, Adamson, and
  Pople]{gillCoulombattenuatedExchangeEnergy1996}
Gill,~P. M.~W.; Adamson,~R.~D.; Pople,~J.~A. Coulomb-Attenuated Exchange Energy
  Density Functionals. \emph{Mol. Phys.} \textbf{1996}, \emph{88},
  1005--1009\relax
\mciteBstWouldAddEndPuncttrue
\mciteSetBstMidEndSepPunct{\mcitedefaultmidpunct}
{\mcitedefaultendpunct}{\mcitedefaultseppunct}\relax
\EndOfBibitem
\bibitem[Leininger \latin{et~al.}(1997)Leininger, Stoll, Werner, and
  Savin]{leiningerCombiningLongrangeConfiguration1997a}
Leininger,~T.; Stoll,~H.; Werner,~H.-J.; Savin,~A. Combining Long-Range
  Configuration Interaction with Short-Range Density Functionals. \emph{Chem.
  Phys. Lett.} \textbf{1997}, \emph{275}, 151--160\relax
\mciteBstWouldAddEndPuncttrue
\mciteSetBstMidEndSepPunct{\mcitedefaultmidpunct}
{\mcitedefaultendpunct}{\mcitedefaultseppunct}\relax
\EndOfBibitem
\bibitem[Iikura \latin{et~al.}(2001)Iikura, Tsuneda, Yanai, and
  Hirao]{iikuraLongrangeCorrectionScheme2001}
Iikura,~H.; Tsuneda,~T.; Yanai,~T.; Hirao,~K. A Long-Range Correction Scheme
  for Generalized-Gradient-Approximation Exchange Functionals. \emph{J. Chem.
  Phys.} \textbf{2001}, \emph{115}, 3540--3544\relax
\mciteBstWouldAddEndPuncttrue
\mciteSetBstMidEndSepPunct{\mcitedefaultmidpunct}
{\mcitedefaultendpunct}{\mcitedefaultseppunct}\relax
\EndOfBibitem
\bibitem[Toulouse \latin{et~al.}(2004)Toulouse, Colonna, and
  Savin]{Toulouse04062505}
Toulouse,~J.; Colonna,~F.; Savin,~A. Long-Range-Short-Range Separation of the
  Electron-Electron Interaction in Density-Functional Theory. \emph{Phys. Rev.
  A} \textbf{2004}, \emph{70}, 062505\relax
\mciteBstWouldAddEndPuncttrue
\mciteSetBstMidEndSepPunct{\mcitedefaultmidpunct}
{\mcitedefaultendpunct}{\mcitedefaultseppunct}\relax
\EndOfBibitem
\bibitem[Yanai \latin{et~al.}(2004)Yanai, Tew, and
  Handy]{yanaiNewHybridExchange2004a}
Yanai,~T.; Tew,~D.~P.; Handy,~N.~C. A New Hybrid Exchange\textendash
  Correlation Functional Using the {{Coulomb}}-Attenuating Method
  ({{CAM}}-{{B3LYP}}). \emph{Chem. Phys. Lett.} \textbf{2004}, \emph{393},
  51--57\relax
\mciteBstWouldAddEndPuncttrue
\mciteSetBstMidEndSepPunct{\mcitedefaultmidpunct}
{\mcitedefaultendpunct}{\mcitedefaultseppunct}\relax
\EndOfBibitem
\bibitem[Baer and Neuhauser(2005)Baer, and
  Neuhauser]{baerDensityFunctionalTheory2005}
Baer,~R.; Neuhauser,~D. Density {{Functional Theory}} with {{Correct
  Long}}-{{Range Asymptotic Behavior}}. \emph{Phys. Rev. Lett.} \textbf{2005},
  \emph{94}, 043002\relax
\mciteBstWouldAddEndPuncttrue
\mciteSetBstMidEndSepPunct{\mcitedefaultmidpunct}
{\mcitedefaultendpunct}{\mcitedefaultseppunct}\relax
\EndOfBibitem
\bibitem[Vydrov and Scuseria(2006)Vydrov, and
  Scuseria]{vydrovAssessmentLongrangeCorrected2006}
Vydrov,~O.~A.; Scuseria,~G.~E. Assessment of a Long-Range Corrected Hybrid
  Functional. \emph{J. Chem. Phys.} \textbf{2006}, \emph{125}, 234109\relax
\mciteBstWouldAddEndPuncttrue
\mciteSetBstMidEndSepPunct{\mcitedefaultmidpunct}
{\mcitedefaultendpunct}{\mcitedefaultseppunct}\relax
\EndOfBibitem
\bibitem[Vydrov \latin{et~al.}(2006)Vydrov, Heyd, Krukau, and
  Scuseria]{vydrovImportanceShortrangeLongrange2006}
Vydrov,~O.~A.; Heyd,~J.; Krukau,~A.~V.; Scuseria,~G.~E. Importance of
  Short-Range versus Long-Range {{Hartree}}-{{Fock}} Exchange for the
  Performance of Hybrid Density Functionals. \emph{J. Chem. Phys.}
  \textbf{2006}, \emph{125}, 074106\relax
\mciteBstWouldAddEndPuncttrue
\mciteSetBstMidEndSepPunct{\mcitedefaultmidpunct}
{\mcitedefaultendpunct}{\mcitedefaultseppunct}\relax
\EndOfBibitem
\bibitem[Cohen \latin{et~al.}(2007)Cohen, {Mori-S{\'a}nchez}, and
  Yang]{cohenDevelopmentExchangecorrelationFunctionals2007}
Cohen,~A.~J.; {Mori-S{\'a}nchez},~P.; Yang,~W. Development of
  Exchange-Correlation Functionals with Minimal Many-Electron Self-Interaction
  Error. \emph{J. Chem. Phys.} \textbf{2007}, \emph{126}, 191109\relax
\mciteBstWouldAddEndPuncttrue
\mciteSetBstMidEndSepPunct{\mcitedefaultmidpunct}
{\mcitedefaultendpunct}{\mcitedefaultseppunct}\relax
\EndOfBibitem
\bibitem[Chai and Head-Gordon(2008)Chai, and Head-Gordon]{Chai08084106}
Chai,~J.-D.; Head-Gordon,~M. Systematic Optimization of Long-Range Corrected
  Hybrid Density Functionals. \emph{J. Chem. Phys.} \textbf{2008}, \emph{128},
  084106\relax
\mciteBstWouldAddEndPuncttrue
\mciteSetBstMidEndSepPunct{\mcitedefaultmidpunct}
{\mcitedefaultendpunct}{\mcitedefaultseppunct}\relax
\EndOfBibitem
\bibitem[Baer \latin{et~al.}(2010)Baer, Livshits, and
  Salzner]{baerTunedRangeSeparatedHybrids2010}
Baer,~R.; Livshits,~E.; Salzner,~U. Tuned {{Range}}-{{Separated Hybrids}} in
  {{Density Functional Theory}}. \emph{Annu. Rev. Phys. Chem.} \textbf{2010},
  \emph{61}, 85--109\relax
\mciteBstWouldAddEndPuncttrue
\mciteSetBstMidEndSepPunct{\mcitedefaultmidpunct}
{\mcitedefaultendpunct}{\mcitedefaultseppunct}\relax
\EndOfBibitem
\bibitem[Zhao \latin{et~al.}(2004)Zhao, Lynch, and
  Truhlar]{zhaoDoublyHybridMeta2004}
Zhao,~Y.; Lynch,~B.~J.; Truhlar,~D.~G. Doubly {{Hybrid Meta DFT}}:\, {{New
  Multi}}-{{Coefficient Correlation}} and {{Density Functional Methods}} for
  {{Thermochemistry}} and {{Thermochemical Kinetics}}. \emph{J. Phys. Chem. A}
  \textbf{2004}, \emph{108}, 4786--4791\relax
\mciteBstWouldAddEndPuncttrue
\mciteSetBstMidEndSepPunct{\mcitedefaultmidpunct}
{\mcitedefaultendpunct}{\mcitedefaultseppunct}\relax
\EndOfBibitem
\bibitem[Grimme(2006)]{grimmeSemiempiricalHybridDensity2006}
Grimme,~S. Semiempirical Hybrid Density Functional with Perturbative
  Second-Order Correlation. \emph{J. Chem. Phys.} \textbf{2006}, \emph{124},
  034108\relax
\mciteBstWouldAddEndPuncttrue
\mciteSetBstMidEndSepPunct{\mcitedefaultmidpunct}
{\mcitedefaultendpunct}{\mcitedefaultseppunct}\relax
\EndOfBibitem
\bibitem[Chai and {Head-Gordon}(2009)Chai, and
  {Head-Gordon}]{chaiLongrangeCorrectedDoublehybrid2009}
Chai,~J.-D.; {Head-Gordon},~M. Long-Range Corrected Double-Hybrid Density
  Functionals. \emph{J. Chem. Phys.} \textbf{2009}, \emph{131}, 174105\relax
\mciteBstWouldAddEndPuncttrue
\mciteSetBstMidEndSepPunct{\mcitedefaultmidpunct}
{\mcitedefaultendpunct}{\mcitedefaultseppunct}\relax
\EndOfBibitem
\bibitem[Zhang \latin{et~al.}(2009)Zhang, Xu, and
  Goddard]{zhangDoublyHybridDensity2009}
Zhang,~Y.; Xu,~X.; Goddard,~W.~A. Doubly Hybrid Density Functional for Accurate
  Descriptions of Nonbond Interactions, Thermochemistry, and Thermochemical
  Kinetics. \emph{PNAS} \textbf{2009}, \emph{106}, 4963--4968\relax
\mciteBstWouldAddEndPuncttrue
\mciteSetBstMidEndSepPunct{\mcitedefaultmidpunct}
{\mcitedefaultendpunct}{\mcitedefaultseppunct}\relax
\EndOfBibitem
\bibitem[Su \latin{et~al.}(2014)Su, Yang, {Mori-S{\'a}nchez}, and
  Xu]{suFractionalChargeBehavior2014}
Su,~N.~Q.; Yang,~W.; {Mori-S{\'a}nchez},~P.; Xu,~X. Fractional {{Charge
  Behavior}} and {{Band Gap Predictions}} with the {{XYG3 Type}} of {{Doubly
  Hybrid Density Functionals}}. \emph{J. Phys. Chem. A} \textbf{2014},
  \emph{118}, 9201--9211\relax
\mciteBstWouldAddEndPuncttrue
\mciteSetBstMidEndSepPunct{\mcitedefaultmidpunct}
{\mcitedefaultendpunct}{\mcitedefaultseppunct}\relax
\EndOfBibitem
\bibitem[Su \latin{et~al.}(2020)Su, Mahler, and
  Yang]{suPreservingSymmetryDegeneracy2020a}
Su,~N.~Q.; Mahler,~A.; Yang,~W. Preserving {{Symmetry}} and {{Degeneracy}} in
  the {{Localized Orbital Scaling Correction Approach}}. \emph{J. Phys. Chem.
  Lett.} \textbf{2020}, \emph{11}, 1528--1535\relax
\mciteBstWouldAddEndPuncttrue
\mciteSetBstMidEndSepPunct{\mcitedefaultmidpunct}
{\mcitedefaultendpunct}{\mcitedefaultseppunct}\relax
\EndOfBibitem
\bibitem[Mei \latin{et~al.}(2019)Mei, Li, Su, and
  Yang]{meiApproximatingQuasiparticleExcitation2019b}
Mei,~Y.; Li,~C.; Su,~N.~Q.; Yang,~W. Approximating {{Quasiparticle}} and
  {{Excitation Energies}} from {{Ground State Generalized
  Kohn}}\textendash{{Sham Calculations}}. \emph{J. Phys. Chem. A}
  \textbf{2019}, \emph{123}, 666--673\relax
\mciteBstWouldAddEndPuncttrue
\mciteSetBstMidEndSepPunct{\mcitedefaultmidpunct}
{\mcitedefaultendpunct}{\mcitedefaultseppunct}\relax
\EndOfBibitem
\bibitem[SI()]{SI}
see supporting information for more details\relax
\mciteBstWouldAddEndPuncttrue
\mciteSetBstMidEndSepPunct{\mcitedefaultmidpunct}
{\mcitedefaultendpunct}{\mcitedefaultseppunct}\relax
\EndOfBibitem
\bibitem[Becke(1993)]{beckeDensityFunctionalThermochemistry1993a}
Becke,~A.~D. Density-functional Thermochemistry. {{III}}. {{The}} Role of Exact
  Exchange. \emph{J. Chem. Phys.} \textbf{1993}, \emph{98}, 5648--5652\relax
\mciteBstWouldAddEndPuncttrue
\mciteSetBstMidEndSepPunct{\mcitedefaultmidpunct}
{\mcitedefaultendpunct}{\mcitedefaultseppunct}\relax
\EndOfBibitem
\bibitem[Knowles and Werner(1988)Knowles, and
  Werner]{knowlesEfficientMethodEvaluation1988}
Knowles,~P.~J.; Werner,~H.-J. An Efficient Method for the Evaluation of
  Coupling Coefficients in Configuration Interaction Calculations.
  \emph{Chemical Physics Letters} \textbf{1988}, \emph{145}, 514--522\relax
\mciteBstWouldAddEndPuncttrue
\mciteSetBstMidEndSepPunct{\mcitedefaultmidpunct}
{\mcitedefaultendpunct}{\mcitedefaultseppunct}\relax
\EndOfBibitem
\bibitem[Langhoff and Davidson(1974)Langhoff, and
  Davidson]{langhoffConfigurationInteractionCalculations1974}
Langhoff,~S.~R.; Davidson,~E.~R. Configuration Interaction Calculations on the
  Nitrogen Molecule. \emph{Int. J. Quantum Chem.} \textbf{1974}, \emph{8},
  61--72\relax
\mciteBstWouldAddEndPuncttrue
\mciteSetBstMidEndSepPunct{\mcitedefaultmidpunct}
{\mcitedefaultendpunct}{\mcitedefaultseppunct}\relax
\EndOfBibitem
\bibitem[Werner and Knowles(1988)Werner, and
  Knowles]{wernerEfficientInternallyContracted1988}
Werner,~H.-J.; Knowles,~P.~J. An Efficient Internally Contracted
  Multiconfiguration\textendash Reference Configuration Interaction Method.
  \emph{J. Chem. Phys.} \textbf{1988}, \emph{89}, 5803--5814\relax
\mciteBstWouldAddEndPuncttrue
\mciteSetBstMidEndSepPunct{\mcitedefaultmidpunct}
{\mcitedefaultendpunct}{\mcitedefaultseppunct}\relax
\EndOfBibitem
\bibitem[Mulliken(1955)]{mullikenElectronicPopulationAnalysis1955}
Mulliken,~R.~S. Electronic {{Population Analysis}} on {{LCAO}}\textendash{{MO
  Molecular Wave Functions}}. {{I}}. \emph{J. Chem. Phys.} \textbf{1955},
  \emph{23}, 1833--1840\relax
\mciteBstWouldAddEndPuncttrue
\mciteSetBstMidEndSepPunct{\mcitedefaultmidpunct}
{\mcitedefaultendpunct}{\mcitedefaultseppunct}\relax
\EndOfBibitem
\bibitem[Grimme \latin{et~al.}(2011)Grimme, Ehrlich, and
  Goerigk]{grimmeEffectDampingFunction2011}
Grimme,~S.; Ehrlich,~S.; Goerigk,~L. Effect of the Damping Function in
  Dispersion Corrected Density Functional Theory. \emph{J. Comput. Chem.}
  \textbf{2011}, \emph{32}, 1456--1465\relax
\mciteBstWouldAddEndPuncttrue
\mciteSetBstMidEndSepPunct{\mcitedefaultmidpunct}
{\mcitedefaultendpunct}{\mcitedefaultseppunct}\relax
\EndOfBibitem
\bibitem[M{\o}ller and Plesset(1934)M{\o}ller, and
  Plesset]{mollerNoteApproximationTreatment1934}
M{\o}ller,~C.; Plesset,~M.~S. Note on an {{Approximation Treatment}} for
  {{Many}}-{{Electron Systems}}. \emph{Phys. Rev.} \textbf{1934}, \emph{46},
  618--622\relax
\mciteBstWouldAddEndPuncttrue
\mciteSetBstMidEndSepPunct{\mcitedefaultmidpunct}
{\mcitedefaultendpunct}{\mcitedefaultseppunct}\relax
\EndOfBibitem
\bibitem[Streets \latin{et~al.}(1972)Streets, Elane~Hall, and
  Ceasar]{streetsMesomericMixingEnergy1972}
Streets,~D.~G.; Elane~Hall,~W.; Ceasar,~G.~P. Mesomeric Mixing in the {$\pi$}
  Energy Levels of Amino-Benzenes Studied by Photoelectron Spectroscopy.
  \emph{Chem. Phys. Lett.} \textbf{1972}, \emph{17}, 90--94\relax
\mciteBstWouldAddEndPuncttrue
\mciteSetBstMidEndSepPunct{\mcitedefaultmidpunct}
{\mcitedefaultendpunct}{\mcitedefaultseppunct}\relax
\EndOfBibitem
\bibitem[Khuseynov \latin{et~al.}(2012)Khuseynov, Fontana, and
  Sanov]{khuseynovPhotoelectronSpectroscopyPhotochemistry2012}
Khuseynov,~D.; Fontana,~M.~T.; Sanov,~A. Photoelectron Spectroscopy and
  Photochemistry of Tetracyanoethylene Radical Anion in the Gas Phase.
  \emph{Chem. Phys. Lett.} \textbf{2012}, \emph{550}, 15--18\relax
\mciteBstWouldAddEndPuncttrue
\mciteSetBstMidEndSepPunct{\mcitedefaultmidpunct}
{\mcitedefaultendpunct}{\mcitedefaultseppunct}\relax
\EndOfBibitem
\bibitem[Cohen \latin{et~al.}(2008)Cohen, Mori-S{\'a}nchez, and Yang]{Cohen08}
Cohen,~A.~J.; Mori-S{\'a}nchez,~P.; Yang,~W. Fractional charge perspective on
  the band gap in density-functional theory. \emph{Phys. Rev. B} \textbf{2008},
  \emph{77}\relax
\mciteBstWouldAddEndPuncttrue
\mciteSetBstMidEndSepPunct{\mcitedefaultmidpunct}
{\mcitedefaultendpunct}{\mcitedefaultseppunct}\relax
\EndOfBibitem
\end{mcitethebibliography}
 
\end{document}


\section{I. Supplemental results for SCF-LOSC}
\subsection{1. Molecular dissociation}
If not specified, below are the calculation details for cases of
molecular dissociation studied in this work.
All the DFT calculation are performed from using an in-house developed
$\rm{QM}^4\rm{D}$ package \cite{qm4d}, if not specified.
The MRCI+Q calculation is performed from using GAMESS package.
\cite{gordonChapter41Advances2005,schmidtGeneralAtomicMolecular1993}
The GW calculation is performed from using FHI-aims package.
\cite{blumInitioMolecularSimulations2009,renResolutionofidentityApproachHartree2012}
The MP2 and the D3 version of Grimme's dispersion with Becke-Johnson
damping (D3BJ) are performed from using Gaussian 16 package \cite{g16}.
cc-pVTZ is used as the basis set for all the calculations.
For LOSC calculations, aug-cc-pVTZ-RIFIT is used as the fitting basis set
in the construction of LOSC curvature matrix.
The molecular geometries for 1,4-benzenediamine and tetracyanoethylene (TCNE)
are optimized from using B3LYP/6-31g* with Gaussian 16 package.
\cite{g16}
In the following text, the D-A complex refers to the
1,4-benzenediamine/TCNE complex system. The donor molecule
refers to 1,4-benzenediamine and the acceptor refers to
TCNE.

\begin{figure}[htbp]
    \centering
    \subfigure[Mulliken charge]
        {\includegraphics[width=0.48\linewidth]{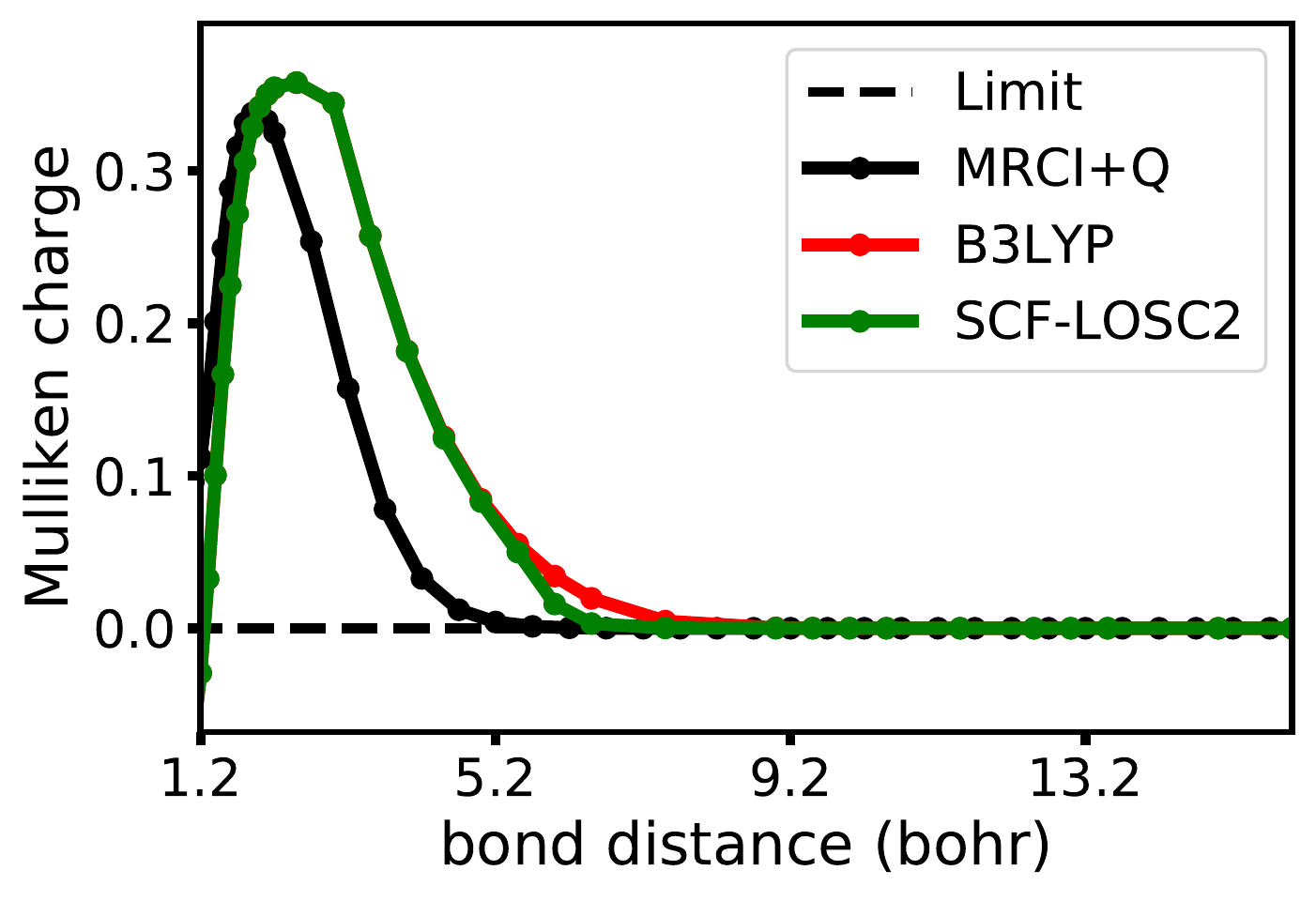}}
    \subfigure[Relative total energy]
        {\includegraphics[width=0.48\linewidth]{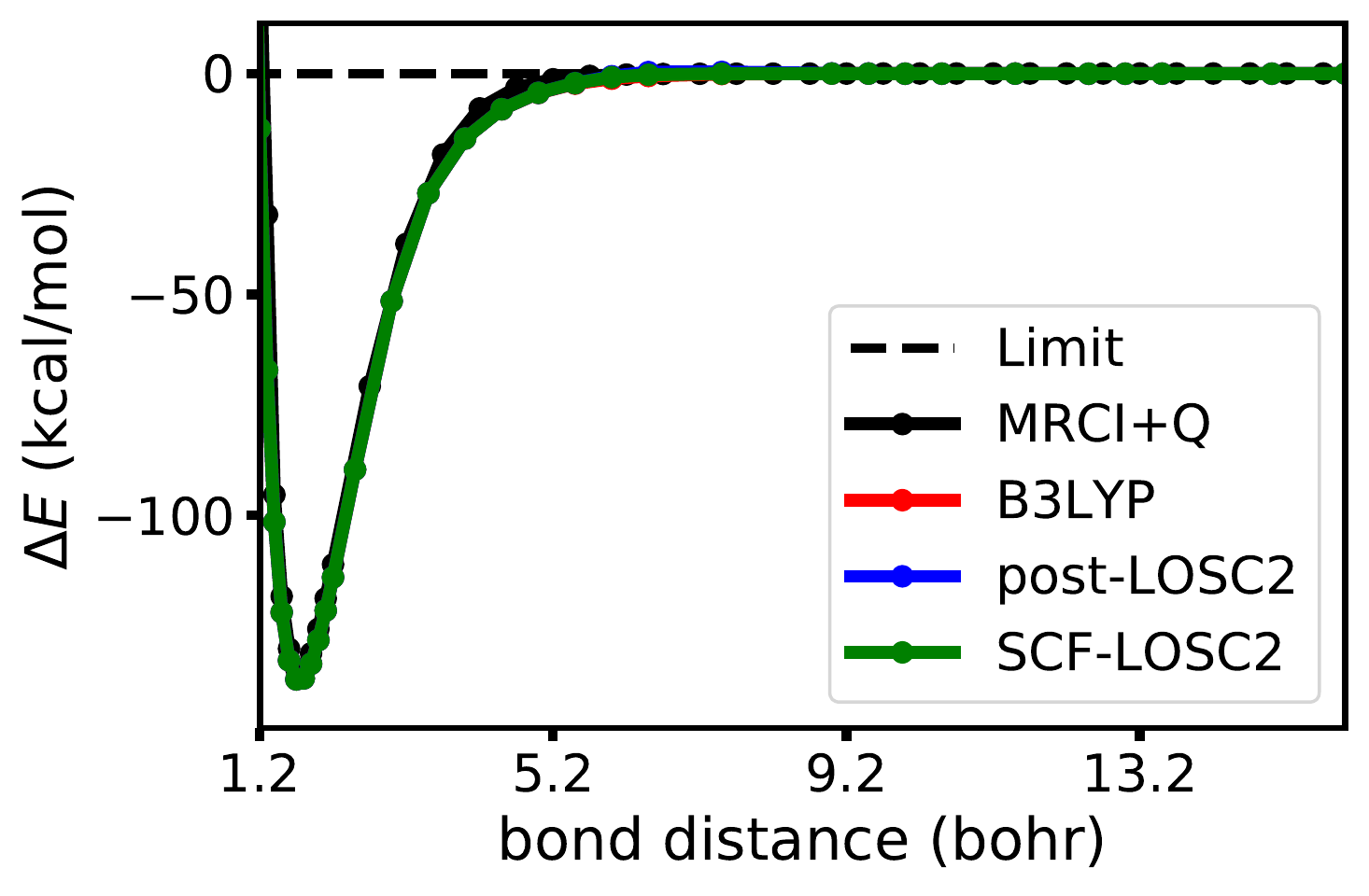}}
    \caption{Dissociation of HF molecule: (a) the Mulliken charge of H atom; (b)
        the relative total energy of HF with respect to H and F atoms.
        $\Delta E = E_{\rm{HF}} - (E_{\rm{H}} + E_{\rm{F}})$. The LOSC2
        calculation is associated with B3LYP.}
    \label{fig:HF}
\end{figure}

Figure \ref{fig:HF} shows the dissociation results for HF molecule.
Since $\rm{IP}_{\rm{H}} > \rm{EA}_{\rm{F}}$, the HF molecule must dissociate
into neutral H and F atom. According to Figrue \ref{fig:HF},
we see that B3LYP already describes the electron density quite well
for this system. In addition, we see SCF-LOSC-B3LYP
preserves the good results for the Mulliken charge, and
both SCF-LOSC-B3LYP and post-LOSC-B3LYP preserve the good results for the
relative total energy.

Tables \ref{tab:HF_E} -- \ref{tab:LiF_mk} show the detailed data
for the relative total energies and Mulliken charges for the HF,
LiH and LiF molecules along their dissociations. Tables \ref{tab:DA_E}
-- \ref{tab:DA_ea} show the detailed data for the relative
total energies, Mulliken charges, first ionization
potentials (IPs) and electron affinities (EAs) for the D-A complex
system along its dissociation.

\newpage
\renewcommand\arraystretch{0.6}
\begin{longtable}{crrrr}
    \caption{Relative total energies (in kcal/mol) of HF molecule to H and F atoms along
            the dissociation of HF. LOSC2 calculation is associated with
            B3LYP. The bond length is in bohr.}
    \label{tab:HF_E} \\
    \toprule
    bond length & B3LYP & post-LOSC2 & SCF-LOSC2 & MRCI+Q   \\ \midrule
    \endfirsthead
    \multicolumn{5}{l}{Table \ref{tab:HF_E} continued: HF relative total energy.} \\ \toprule
    bond length & B3LYP & post-LOSC2 & SCF-LOSC2 & MRCI+Q   \\ \midrule
    \endhead
    \midrule
    \endfoot
    \endlastfoot
1.2  & -12.4575  & -12.4574   & -12.4574  & 30.2242   \\
1.3  & -67.1318  & -67.1317   & -67.1317  & -31.9542  \\
1.4  & -101.5012 & -101.5012  & -101.5012 & -95.3494  \\
1.5  & -121.9666 & -121.9666  & -121.9666 & -118.3303 \\
1.6  & -132.8496 & -132.8496  & -132.8495 & -130.1832 \\
1.7  & -137.1426 & -137.1425  & -137.1425 & -134.8318 \\
1.8  & -136.9154 & -136.9153  & -136.9153 & -134.659  \\
1.9  & -133.6136 & -133.6134  & -133.6134 & -131.2544 \\
\bottomrule
\end{longtable}

\newpage
\renewcommand\arraystretch{0.6}
\begin{longtable}{crrr}
    \caption{Mulliken charges of H atom along the dissociation of HF molecule.
            LOSC2 calculation is associated with B3LYP.
            Bond length is in bohr.}
    \label{tab:HF_mk}\\
    \toprule
    bond length & B3LYP   & SCF-LOSC2 & MRCI+Q \\ \midrule
    \endfirsthead
    \multicolumn{4}{l}{Table \ref{tab:HF_E} continued: Mulliken charge for HF.} \\ \toprule
    bond length & B3LYP   & SCF-LOSC2 & MRCI+Q \\ \midrule
    \endhead
    \midrule
    \endfoot
    \endlastfoot
    1.2  & -0.0294 & -0.0293   & 0.1115 \\
    1.3  & 0.0324  & 0.0324    & 0.1518 \\
    1.4  & 0.1004  & 0.1004    & 0.2011 \\
    1.5  & 0.1664  & 0.1664    & 0.2489 \\
    1.6  & 0.2249  & 0.2249    & 0.2882 \\
    1.7  & 0.2718  & 0.2719    & 0.3156 \\
    1.8  & 0.3058  & 0.3058    & 0.3314 \\
    1.9  & 0.3281  & 0.3282    & 0.3379 \\
    2.0  & 0.3418  & 0.3419    & 0.3376 \\
    2.1  & 0.3498  & 0.3499    & 0.3329 \\
    2.2  & 0.3543  & 0.3544    & 0.3250 \\
    2.5  & 0.3577  & 0.3579    & 0.2875 \\
    3.0  & 0.3444  & 0.3444    & 0.1959 \\
    3.5  & 0.2574  & 0.2574    & 0.1058 \\
    4.0  & 0.1819  & 0.1816    & 0.0471 \\
    4.5  & 0.1255  & 0.1245    & 0.0182 \\
    5.0  & 0.0846  & 0.0831    & 0.0064 \\
    5.5  & 0.0551  & 0.0499    & 0.0020 \\
    6.0  & 0.0342  & 0.0159    & 0.0006 \\
    6.5  & 0.0196  & 0.0031    & 0.0001 \\
    7.5  & 0.0047  & 0.0002    & 0.0000 \\
    9.0  & 0.0003  & 0.0000    & 0.0000 \\
    9.5  & 0.0001  & 0.0000    & 0.0000 \\
    10.0 & 0.0000  & 0.0000    & 0.0000 \\
    10.5 & 0.0000  & 0.0000    & 0.0000 \\
    11.5 & 0.0000  & 0.0000    & 0.0000 \\
    12.5 & 0.0000  & 0.0000    & 0.0000 \\
    13.0 & 0.0000  & 0.0000    & 0.0000 \\
    13.5 & 0.0000  & 0.0000    & 0.0000 \\
    15.0 & 0.0000  & 0.0000    & 0.0000 \\
    16.0 & 0.0000  & 0.0000    & 0.0000 \\ \bottomrule
\end{longtable}

\newpage
\renewcommand\arraystretch{0.6}
\begin{longtable}{crrrr}
    \caption{Relative total energies (in kcal/mol) of LiH molecule to Li and H
            atoms along the dissociation of LiH. LOSC2 calculation is associated
            with B3LYP. Bond length is in bohr.}
    \label{tab:LiH_E}\\
    \toprule
    bond length & B3LYP & post-LOSC2 & SCF-LOSC2 & MRCI+Q   \\ \midrule
    \endfirsthead
    \multicolumn{5}{l}{Table \ref{tab:LiH_E} continued: LiH relative total energy.} \\ \toprule
    bond length & B3LYP & post-LOSC2 & SCF-LOSC2 & MRCI+Q   \\ \midrule
    \endhead
    \midrule
    \endfoot
    \endlastfoot
    2.0  & -15.4288 & -15.3627   & -15.3710  & -10.7844 \\
    2.1  & -26.1238 & -26.0563   & -26.0645  & -21.9963 \\
    2.2  & -34.6988 & -34.6291   & -34.6374  & -30.9907 \\
    2.3  & -41.4997 & -41.4273   & -41.4357  & -38.1313 \\
    2.4  & -46.8077 & -46.7320   & -46.7406  & -43.7218 \\
    2.5  & -50.8853 & -50.8059   & -50.8147  & -48.0160 \\
    2.6  & -53.9230 & -53.8392   & -53.8483  & -51.2262 \\
    2.7  & -56.0525 & -55.9640   & -55.9733  & -53.5307 \\
\bottomrule
\end{longtable}

\renewcommand\arraystretch{0.6}
\begin{longtable}{crrr}
    \caption{Mulliken charges of Li atom along the dissociation of LiH molecule.
            LOSC2 calculation is associated with B3LYP.
            Bond length is in bohr.}
    \label{tab:LiH_mk}\\
    \toprule
    bond length & B3LYP   & SCF-LOSC2 & MRCI+Q \\ \midrule
    \endfirsthead
    \multicolumn{4}{l}{Table \ref{tab:LiH_mk} continued: Mulliken charge for LiH.} \\ \toprule
    bond length & B3LYP   & SCF-LOSC2 & MRCI+Q \\ \midrule
    \endhead
    \midrule
    \endfoot
    \endlastfoot
2.0  & -0.1833 & -0.1845   & 0.0117  \\
2.1  & -0.1026 & -0.1033   & 0.0341  \\
2.2  & -0.0294 & -0.0296   & 0.0558  \\
2.3  & 0.0346  & 0.0347    & 0.0765  \\
2.4  & 0.0887  & 0.0892    & 0.0958  \\
2.5  & 0.1332  & 0.1341    & 0.1138  \\
2.6  & 0.1691  & 0.1705    & 0.1305  \\
2.7  & 0.1980  & 0.1999    & 0.1461  \\
\bottomrule
\end{longtable}

\newpage
\renewcommand\arraystretch{0.6}
\begin{longtable}{crrrr}
    \caption{Relative total energies (in kcal/mol) of LiF molecule to Li and F
            atoms along the dissociation of LiF molecule. LOSC2 calculation is associated
            with B3LYP. Bond length is in bohr.}
    \label{tab:LiF_E}\\
    \toprule
    bond length & B3LYP & post-LOSC2 & SCF-LOSC2 & MRCI+Q   \\ \midrule
    \endfirsthead
    \multicolumn{5}{l}{Table \ref{tab:LiF_E} continued: LiF relative total energy:} \\ \toprule
    bond length & B3LYP & post-LOSC2 & SCF-LOSC2 & MRCI+Q   \\ \midrule
    \endhead
    \midrule
    \endfoot
    \endlastfoot
2.0  & -11.1313  & -11.1282   & -11.1292  & 7.6169    \\
2.1  & -46.2190  & -46.2168   & -46.2176  & -32.2749  \\
2.2  & -72.8208  & -72.8190   & -72.8196  & -62.3335  \\
2.3  & -93.5515  & -93.5499   & -93.5505  & -84.7549  \\
2.4  & -108.7332 & -108.7308  & -108.7313 & -101.2518 \\
2.5  & -119.4273 & -119.4239  & -119.4246 & -113.1557 \\
2.6  & -127.0197 & -127.0153  & -127.0161 & -121.5001 \\
2.7  & -131.9696 & -131.9637  & -131.9648 & -127.0870 \\
\bottomrule
\end{longtable}

\renewcommand\arraystretch{0.6}
\begin{longtable}{crrr}
    \caption{Mulliken charges of Li atom along the dissociation of LiF molecule.
            LOSC2 calculation is associated with B3LYP.
            Bond length is in bohr.}
    \label{tab:LiF_mk}\\
    \toprule
    bond length & B3LYP   & SCF-LOSC2 & MRCI+Q \\ \midrule
    \endfirsthead
    \multicolumn{4}{l}{Table \ref{tab:LiF_mk} continued: Mulliken charge for LiF.} \\ \toprule
    bond length & B3LYP   & SCF-LOSC2 & MRCI+Q \\ \midrule
    \endhead
    \midrule
    \endfoot
    \endlastfoot
2.0  & 0.1009 & 0.1016    & 0.4072 \\
2.1  & 0.1382 & 0.1389    & 0.4185 \\
2.2  & 0.1698 & 0.1696    & 0.4293 \\
2.3  & 0.1963 & 0.1970    & 0.4398 \\
2.4  & 0.2236 & 0.2245    & 0.4502 \\
2.5  & 0.2528 & 0.2537    & 0.4605 \\
2.6  & 0.2838 & 0.2848    & 0.4707 \\
2.7  & 0.3164 & 0.3175    & 0.4807 \\
\bottomrule
\end{longtable}

\newpage
\newgeometry{left=0.8in, right=0.5in, top=1in, bottom=1in}
\begin{landscape}
\renewcommand\arraystretch{0.6}
\begin{longtable}{crrrrrrr}
    \caption{Relative total energies (in kcal/mol) of the D-A complex system to
            the donor and acceptor subsystems along the dissociation of
            D-A complex system.
            LOSC2 calculation is associated with BLYP.
            The separation distance of D-A complex is in Angstrom.
            D3BJ is obtained from BLYP and added to all the DFT
            energy results.}
    \label{tab:DA_E}\\
    \toprule
    distance & MP2      & BLYP    & post-LOSC2 & SCF-LOSC2 & BLYP+D3BJ & post-LOSC2+D3BJ & SCF-LOSC2+D3BJ \\ \midrule
    \endfirsthead
    \multicolumn{7}{l}{Table \ref{tab:DA_E} continued: D-A relative total energy:} \\ \toprule
    distance & MP2      & BLYP    & post-LOSC2 & SCF-LOSC2 & BLYP+D3BJ & post-LOSC2+D3BJ & SCF-LOSC2+D3BJ \\ \midrule
    \endhead
    \midrule
    \endfoot
    \endlastfoot
    2.0      & -10.1188 & 24.0480 & 23.9289    & 23.9311   & 0.2372    & 0.1180          & 0.1202         \\
    2.1      & -14.7578 & 15.5839 & 15.5148    & 15.5121   & -6.7456   & -6.8147         & -6.8174        \\
    2.2      & -17.4782 & 9.4401  & 9.4186     & 9.4096    & -11.3526  & -11.3741        & -11.3831       \\
    2.3      & -18.8295 & 5.1131  & 5.1393     & 5.1237    & -14.1209  & -14.0947        & -14.1103       \\
    2.4      & -19.2206 & 2.1510  & 2.2190     & 2.2137    & -15.5295  & -15.4614        & -15.4668       \\
    2.5      & -18.9537 & -0.0009 & 0.1769     & 0.2277    & -16.1564  & -15.9785        & -15.9277       \\
    2.6      & -18.2509 & -1.5905 & -1.3902    & -1.1377   & -16.2699  & -16.0696        & -15.8172       \\
    2.7      & -17.2745 & -2.7724 & -2.5995    & -2.0587   & -16.0423  & -15.8693        & -15.3285       \\
\bottomrule
\end{longtable}
\end{landscape}
\restoregeometry

\renewcommand\arraystretch{0.6}
\begin{longtable}{crrr}
    \caption{Mulliken charges of the donor molecule along the dissociation
            of D-A complex system.
            LOSC2 calculation is associated with B3LYP.
            Separation distance is in Angstrom.}
    \label{tab:DA_mk}\\
    \toprule
    distance & MP2    & BLYP   & SCF-LOSC2 \\ \midrule
    \endfirsthead
    distance & MP2    & BLYP   & SCF-LOSC2 \\ \midrule
    \endhead
    \midrule
    \endfoot
    \endlastfoot
2.0  & 0.2060 & 0.2675 & 0.2666    \\
2.1  & 0.1678 & 0.2336 & 0.2331    \\
2.2  & 0.1358 & 0.2031 & 0.2025    \\
2.3  & 0.1097 & 0.1756 & 0.1738    \\
2.4  & 0.0887 & 0.1773 & 0.1478    \\
2.5  & 0.0720 & 0.1829 & 0.1246    \\
2.6  & 0.0586 & 0.1923 & 0.1035    \\
2.7  & 0.0479 & 0.1993 & 0.0844    \\
2.8  & 0.0393 & 0.2043 & 0.0675    \\
2.9  & 0.0323 & 0.2082 & 0.0534    \\
3.0  & 0.0266 & 0.2115 & 0.0419    \\
3.1  & 0.0219 & 0.2141 & 0.0326    \\
3.2  & 0.0181 & 0.2164 & 0.0252    \\
3.3  & 0.0150 & 0.2185 & 0.0194    \\
3.4  & 0.0124 & 0.2200 & 0.0152    \\
3.5  & 0.0102 & 0.2212 & 0.0121    \\
3.6  & 0.0085 & 0.2222 & 0.0100    \\
3.7  & 0.0070 & 0.2228 & 0.0083    \\
3.8  & 0.0058 & 0.2230 & 0.0070    \\
3.9  & 0.0047 & 0.2231 & 0.0059    \\
4.0  & 0.0039 & 0.2232 & 0.0050    \\
5.0  & 0.0003 & 0.2190 & 0.0005    \\
6.0  & 0.0000 & 0.2128 & 0.0000    \\
7.0  & 0.0000 & 0.2069 & 0.0000    \\
8.0  & 0.0000 & 0.2017 & 0.0000    \\
9.0  & 0.0000 & 0.1971 & 0.0000    \\
10.0 & 0.0000 & 0.1932 & 0.0000    \\ \bottomrule
\end{longtable}

\newpage
\renewcommand\arraystretch{0.6}
\begin{longtable}{crrrrrr}
    \caption{The first ionization potentials (IPs, in eV) of the D-A complex along the
            its dissociation. The negative energy of the highest occupied molecular
            orbital from DFT is used to evaluate the IP. The IP of the
            donor molecule is also listed in the table for comparison.
            LOSC2 and GW calculation is associated with BLYP.
            evGW refers to eigenvalue self-consistent GW calculation.
            Distance is in Angstrom.}
    \label{tab:DA_ip}\\
    \toprule
    distance & G$_0$W$_0$ & evGW   & BLYP   & post-LOSC2 & SCF-LOSC2 \\ \midrule
    \endfirsthead
    distance & G$_0$W$_0$ & evGW   & BLYP   & post-LOSC2 & SCF-LOSC2 \\ \midrule
    \endhead
    \midrule
    \endfoot
    \endlastfoot
    2.0   & 7.1655 & 7.8547 & 5.0565 & 8.2385     & 8.2359    \\
    3.0   & 6.7645 & 7.1875 & 4.9001 & 8.0881     & 7.6724    \\
    4.0   & 6.7213 & 7.1155 & 4.8926 & 8.0395     & 7.4042    \\
    5.0   & 6.7431 & 7.1306 & 4.8833 & 8.0136     & 7.2949    \\
    6.0   & 6.7608 & 7.1494 & 4.8755 & 7.9978     & 7.2351    \\
    7.0   & 6.7752 & 7.1650 & 4.8695 & 7.9882     & 7.1993    \\
    8.0   & 6.7885 & 7.1734 & 4.8649 & 7.9819     & 7.1762    \\
    9.0   & 6.7996 & 7.1846 & 4.8614 & 7.9775     & 7.1605    \\
    10.0  & 6.8092 & 7.1946 & 4.8586 & 7.9743     & 7.1495    \\ \midrule
    Donor & 6.5812 & 7.0720 & 4.0021 & 7.1168     & 7.1153    \\ \bottomrule
\end{longtable}

\renewcommand\arraystretch{0.6}
\begin{longtable}{crrrrrr}
    \caption{The first electron affinities (EAs, in eV) of D-A complex along
            its dissociation. The negative energy of the lowest unoccupied molecular
            orbital from DFT is used to evaluate the EA. The EA of the
            acceptor molecule is also listed in the table for comparison.
            LOSC2 and GW calculation is associated with BLYP.
            evGW refers to the eigenvalue self-consistent calculation.
            Distance is in Angstrom.}
    \label{tab:DA_ea}\\
    \toprule
    distance & G$_0$W$_0$ & evGW   & BLYP   & post-LOSC2 & SCF-LOSC2 \\ \midrule
    \endfirsthead
    distance & G$_0$W$_0$ & evGW   & BLYP   & post-LOSC2 & SCF-LOSC2 \\ \midrule
    \endhead
    \midrule
    \endfoot
    \endlastfoot
2.0      & 2.5904 & 2.2184 & 4.4780 & 2.3862     & 2.3950    \\
3.0      & 1.7791 & 1.5800 & 4.7850 & 2.7272     & 3.1431    \\
4.0      & 1.8738 & 1.6752 & 4.7620 & 2.7066     & 3.3499    \\
5.0      & 1.8834 & 1.6862 & 4.7539 & 2.7012     & 3.4283    \\
6.0      & 1.8738 & 1.6748 & 4.7495 & 2.6964     & 3.4680    \\
7.0      & 1.8610 & 1.6626 & 4.7471 & 2.6937     & 3.4919    \\
8.0      & 1.8476 & 1.6486 & 4.7456 & 2.6919     & 3.5072    \\
9.0      & 1.8359 & 1.6337 & 4.7447 & 2.6907     & 3.5174    \\
10.0     & 1.8255 & 1.6229 & 4.7442 & 2.6899     & 3.5246    \\ \midrule
Acceptor & 3.3117 & 2.9932 & 5.5860 & 3.5161     & 3.5304    \\ \bottomrule
\end{longtable}

\newpage
\newgeometry{left=0.8in, right=0.5in, top=0.6in, bottom=0.7in}
\subsection{2. Atomization energies, reaction barriers, IPs and EAs}

We use the same test sets to investigate the performance of the new
SCF-LOSC approach for the atomization energies, reaction barriers,
IPs and EAs. See Ref \citenum{suPreservingSymmetryDegeneracy2020a}
for more information
about molecular geometries and reference values for these test sets.
Table \ref{tab:G2_RB_eig} shows the results from different methods
for atomization energies, reaction barriers, IPs and EAs.
Detailed data for each test set are summarized in Tables \ref{tab:AE_G2_1} --
\ref{tab:EA}.
The negative HOMO/LUMO energies from DFT calculation is used to evaluate
the first IP/EA respectively.
Table \ref{tab:C2H2} shows the comparison of the negative
HOMO energies of polyacetylene from different methods with the first IP.
All the DFT calculation are performed from using an in-house
developed $\rm{QM}^4\rm{D}$ package. \cite{qm4d}
6-311++G(3df, 3pd) is used as the basis set for the results shown in
Tables \ref{tab:G2_RB_eig}, \ref{tab:AE_G2_1} -- \ref{tab:EA}.
cc-pVDZ is used as the basis set for the results shown in
Table \ref{tab:C2H2}. aug-cc-pVTZ-RIFIT is the fitting basis used
in the construction of curvature matrix in LOSC.

\renewcommand\arraystretch{1.05}
\begin{table}[]
    \caption{Mean absolute errors (MAEs) for atomization energies (AE),
            reaction barriers (RB), IPs and EAs from different methods.
            Results of AE and RB test sets is in kcal/mol.
            Results of IP and EA test sets is in eV.}
    \label{tab:G2_RB_eig}
    \scalebox{0.9}{
\begin{tabular}{@{}l|rrr|rrr|rrr@{}}
    \toprule
                             & \multicolumn{3}{c}{DFA} & \multicolumn{3}{c}{post-LOSC2} & \multicolumn{3}{c}{SCF-LOSC2} \\ \hline
    Test set                 & LDA    & BLYP  & B3LYP & LDA    & BLYP          & B3LYP & LDA    & BLYP         & B3LYP \\ \hline
    \textbf{AE}             &        &       &       &                &                 &                  &               &                &                 \\
    G2-1           & 42.20  & 4.94  & 2.45  & 42.19          & 4.93            & 2.45             & 42.19         & 4.93           & 2.45            \\
    NonHydrocarbon & 97.74  & 14.73 & 7.64  & 97.74          & 14.73           & 7.65             & 97.74         & 14.73          & 7.65            \\
    Hydrocarbon    & 140.95 & 10.36 & 3.47  & 140.94         & 10.38           & 3.48             & 140.94        & 10.38          & 3.48            \\
    SubHydrocarbon & 124.34 & 6.23  & 2.52  & 124.32         & 6.24            & 2.53             & 124.32        & 6.24           & 2.52            \\
    Radical        & 78.02  & 3.89  & 2.26  & 78.00          & 3.91            & 2.26             & 78.00         & 3.91           & 2.26            \\
\bottomrule
    \end{tabular}
    }
\end{table}

\renewcommand\arraystretch{1.05}
\begin{table}[htbp]
    \caption{Negative HOMO energies (in eV) of polyacetylene with different unit numbers
             calculated from different DFAs and compared with the first IP from
             RASPT2.}
    \label{tab:C2H2}
    \scalebox{0.9}{
    \begin{tabular}{lr|rrr|rrr|rrr}
    \toprule
          &        & \multicolumn{3}{c}{DFA} & \multicolumn{3}{c}{post-LOSC2} & \multicolumn{3}{c}{SCF-LOSC2} \\ \hline
    units & RASPT2 & LDA   & BLYP  & B3LYP& LDA      & BLYP     & B3LYP    & LDA      & BLYP     & B3LYP    \\ \hline
    1     & 10.48  & 6.71  & 6.34  & 7.46 & 10.46    & 10.11    & 10.46    & 10.46    & 10.11    & 10.46   \\
    2     & 9.18   & 5.84  & 5.46  & 6.43 & 9.49     & 9.12     & 8.35     & 9.49     & 9.12     & 8.35    \\
    3     & 8.18   & 5.40  & 5.01  & 5.89 & 8.84     & 8.47     & 8.03     & 8.84     & 8.47     & 8.03    \\
    4     & 7.69   & 5.14  & 4.74  & 5.56 & 8.01     & 7.66     & 7.77     & 8.01     & 7.66     & 7.77    \\
    5     & 7.33   & 4.97  & 4.56  & 5.34 & 7.78     & 7.41     & 7.59     & 7.80     & 7.42     & 7.60    \\
    6     & 7.04   & 4.84  & 4.43  & 5.17 & 7.56     & 7.19     & 7.41     & 7.59     & 7.21     & 7.42    \\
    7     & 6.85   & 4.75  & 4.33  & 5.05 & 7.48     & 7.10     & 7.19     & 7.53     & 7.15     & 7.20    \\
    8     & 6.66   & 4.68  & 4.25  & 4.95 & 7.45     & 6.95     & 7.15     & 7.53     & 7.03     & 7.18    \\
    9     & 6.56   & 4.62  & 4.19  & 4.87 & 7.23     & 6.85     & 6.93     & 7.32     & 6.94     & 6.97    \\
    10    & 6.41   & 4.57  & 4.14  & 4.80 & 7.08     & 6.61     & 6.84     & 7.19     & 6.72     & 6.89    \\ \hline
    MAE   &        & 2.49   & 2.89   & 2.09  & 0.50      & 0.20      & 0.33      & 0.54      & 0.24      & 0.35     \\ \bottomrule
    \end{tabular}
}
\end{table}
\restoregeometry

\newgeometry{left=0.8in, right=0.5in, top=0.6in, bottom=0.7in}
\renewcommand\arraystretch{1.05}
\begin{table}[]
\scalebox{0.8}{
    \caption{Detailed data for the atomization energies (in kcal/mol) of the G2-1 test set.
            NA means the data is not available because of the SCF convergence issue
            from DFT.}
    \label{tab:AE_G2_1}
    \begin{tabular}{@{}lr|rrr|rrr|rrr@{}}
    \toprule
        &  & \multicolumn{3}{c}{DFA}
        & \multicolumn{3}{c}{post-LOSC2}
        & \multicolumn{3}{c}{SCF-LOSC2} \\ \hline
    Molecule & Ref & LDA    & BLYP  & B3LYP
    & LDA    & BLYP          & B3LYP
    & LDA    & BLYP         & B3LYP \\ \midrule
    BeH       & 49.8  & 60.26  & 56.91  & 57.89  & 60.24          & 56.89           & 57.88            & 60.24         & 56.89          & 57.88           \\
    CH        & 83.9  & 92.27  & 85.5   & 85.43  & 92.26          & 85.49           & 85.43            & 92.26         & 85.49          & 85.43           \\
    CH2\_3B1  & 190   & NA     & 183.78 & 185.02 & NA             & 183.77          & 185.02           & NA            & 183.77         & 185.02          \\
    CH2\_1A1  & 180.6 & 201.83 & 178.97 & 180.82 & 201.82         & 178.97          & 180.81           & 201.82        & 178.97         & 180.81          \\
    CH3       & 306.6 & 338.78 & 306.75 & 309.79 & 338.77         & 306.74          & 309.78           & 338.77        & 306.74         & 309.78          \\
    CH4       & 419.2 & 461.87 & 416.59 & 420.74 & 461.86         & 416.59          & 420.73           & 461.86        & 416.59         & 420.73          \\
\bottomrule
    \end{tabular}
}
\end{table}
\restoregeometry

\renewcommand\arraystretch{1.1}
\begin{landscape}
\begin{table}[]
    \caption{Detailed data for the atomization energies (in kcal/mol) of the
            Hydrocarbon test set. NA means the data is not available because of
            the SCF convergence issue.}
    \label{tab:AE_Hydrocarbon}
    \begin{tabular}{@{}lr|rrr|rrr|rrr@{}}
    \toprule
        &  & \multicolumn{3}{c}{DFA}
        & \multicolumn{3}{c}{post-LOSC2}
        & \multicolumn{3}{c}{SCF-LOSC2} \\ \hline
    Molecule & Ref & LDA    & BLYP  & B3LYP
    & LDA    & BLYP          & B3LYP
    & LDA    & BLYP         & B3LYP \\ \midrule
    propyne               & 704.1  & 800.64  & 700.95  & 701.88  & 800.64         & 700.94          & 701.88           & 800.64        & 700.94         & 701.88          \\
    allene                & 702.4  & 804.06  & 704.02  & 704.07  & 804.05         & 704.01          & 704.07           & 804.05        & 704.01         & 704.07          \\
    cyclopropene          & 681.9  & 784.06  & 676.58  & 678.5   & 784.06         & 676.57          & 678.5            & 784.06        & 676.57         & 678.5           \\
    propene               & 859.2  & 969.6   & 852.61  & 858.3   & 969.59         & 852.6           & 858.29           & 969.59        & 852.6          & 858.29          \\
    cyclopropane          & 851.8  & 967.91  & 841.9   & 849.6   & 967.9          & 841.89          & 849.6            & 967.9         & 841.89         & 849.6           \\
    propane               & 1004.6 & 1127.23 & 992.43  & 1003.09 & 1127.21        & 992.41          & 1003.08          & 1127.21       & 992.41         & 1003.08         \\
\bottomrule
    \end{tabular}
\end{table}
\end{landscape}

\begin{landscape}
\renewcommand\arraystretch{0.6}
\begin{longtable}{lr|rrr|rrr|rrr}
    \caption{Detailed data for the atomization energies (in kcal/mol) of the
            SubHydrocarbon test set. NA means the data is not available because
            of the SCF convergence issue.}
    \label{tab:AE_SubHydrocarbon}\\
    \toprule
        &  & \multicolumn{3}{c}{DFA}
        & \multicolumn{3}{c}{post-LOSC2}
        & \multicolumn{3}{c}{SCF-LOSC2} \\ \hline
    Molecule & Ref & LDA    & BLYP  & B3LYP
    & LDA    & BLYP          & B3LYP
    & LDA    & BLYP         & B3LYP \\ \midrule
    \endfirsthead
    \multicolumn{11}{l}{Table \ref{tab:AE_SubHydrocarbon} continued.} \\
    \toprule
        &  & \multicolumn{3}{c}{DFA}
        & \multicolumn{3}{c}{post-LOSC2}
        & \multicolumn{3}{c}{SCF-LOSC2} \\ \hline
    Molecule & Ref & LDA    & BLYP  & B3LYP
    & LDA    & BLYP          & B3LYP
    & LDA    & BLYP         & B3LYP \\ \midrule
    \endhead
    \midrule
    \endfoot
    \endlastfoot
    CH2F2             & 436.3  & 517.68  & 438.77  & 435.74  & 517.68         & 438.77          & 435.74           & 517.68        & 438.77         & 435.74          \\
    CHF3              & 457.8  & 560.52  & 460.62  & 454.71  & 560.51         & 460.61          & 454.7            & 560.51        & 460.61         & 454.7           \\
    CH2Cl2            & 369.6  & NA      & 362.63  & 364.34  & NA             & 362.62          & 364.34           & NA            & 362.62         & 364.34          \\
    CHCl3             & 343.3  & NA      & 333     & 333.01  & NA             & 332.98          & 332.99           & NA            & 332.98         & 332.99          \\
    methylamine       & 581.2  & 660.53  & 581.89  & 584.16  & 660.51         & 581.86          & 584.15           & 660.52        & 581.87         & 584.15          \\
    methylcyanide     & 615.1  & 704.62  & 619.59  & 615.22  & 704.61         & 619.58          & 615.22           & 704.61        & 619.58         & 615.22          \\
    nitromethane      & 600.6  & 742.94  & 618.45  & 602.23  & 742.92         & 618.42          & 602.22           & 742.92        & 618.42         & 602.22          \\
    methylnitrite     & 598.3  & 736.06  & 615.75  & 598.96  & 736.06         & 615.74          & 598.95           & 736.06        & 615.74         & 598.95          \\
\bottomrule
\end{longtable}
\end{landscape}

\renewcommand\arraystretch{1.1}
\begin{table}[]
    \caption{Detailed data for the atomization energies (in kcal/mol) of the
            Radical test set. NA means the data is not available because
            of the SCF convergence issue.}
    \label{tab:AE_Radicals}
    \scalebox{0.83}{
    \begin{tabular}{@{}lr|rrr|rrr|rrr@{}}
    \toprule
        &  & \multicolumn{3}{c}{DFA}
        & \multicolumn{3}{c}{post-LOSC2}
        & \multicolumn{3}{c}{SCF-LOSC2} \\ \hline
    Molecule & Ref & LDA    & BLYP  & B3LYP
    & LDA    & BLYP          & B3LYP
    & LDA    & BLYP         & B3LYP \\ \midrule
    H2       & 109.2 & 112.79  & 109.26  & 110.13  & 112.79         & 109.26          & 110.13           & 112.79        & 109.26         & 110.13          \\
    SH       & 86.7  & NA      & NA      & 88.18   & NA             & NA              & 88.18            & NA            & NA             & 88.18           \\
    CCH      & 265.8 & 307.92  & 264.89  & 262.04  & 307.92         & 264.89          & 262.04           & 307.92        & 264.89         & 262.04          \\
    C2H3     & 443.9 & 506.54  & 446.06  & 446.95  & 506.53         & 446.04          & 446.94           & 506.53        & 446.04         & 446.94          \\
    CH3CO    & 580.9 & 676.85  & NA      & 582.54  & 676.82         & NA              & 582.53           & 676.83        & NA             & 582.53          \\
    H2COH    & 408.9 & 477.29  & NA      & 411.13  & 477.28         & NA              & 411.13           & 477.28        & NA             & 411.13          \\
\bottomrule
    \end{tabular}
    }
\end{table}

\begin{table}[]
    \caption{Detailed data for the reaction barriers (in kcal/mol) of the
            HTBH38 test set. NA means the data is not available because
            of the SCF convergence issue.}
    \label{tab:RB_HTBH38}
    \scalebox{0.9}{
    \begin{tabular}{@{}lr|rrr|rrr|rrr@{}}
    \toprule
        &  & \multicolumn{3}{c}{DFA}
        & \multicolumn{3}{c}{post-LOSC2}
        & \multicolumn{3}{c}{SCF-LOSC2} \\ \hline
    case & Ref & LDA    & BLYP  & B3LYP
    & LDA    & BLYP          & B3LYP
    & LDA    & BLYP         & B3LYP \\ \midrule
\multicolumn{11}{l}{Forward: $E_{\rm{TS}}$ - $\sum E_{\rm{reactant}}$} \\ \hline
TS1  & 5.7   & -3.31  & -2.5   & -0.81 & -3.31          & -2.5            & -0.8             & -3.31         & -2.5           & -0.8            \\
TS2  & 4.9   & -18.38 & -3.49  & 0.52  & -18.38         & -3.49           & 0.52             & -18.38        & -3.49          & 0.52            \\
TS3  & 12.1  & -5.39  & 7.1    & 8.7   & -5.38          & 7.11            & 8.7              & -5.38         & 7.11           & 8.7             \\
TS4  & 6.5   & -17.13 & -2.71  & 1.98  & -17.12         & -2.71           & 1.98             & -17.12        & -2.71          & 1.98            \\
TS5  & 9.6   & -2.88  & 2.86   & 4.22  & -2.88          & 2.86            & 4.22             & -2.88         & 2.86           & 4.22            \\
\bottomrule
\end{tabular}
    }
\end{table}

\begin{table}[]
    \caption{Detailed data for the reaction barriers (in kcal/mol) of the
            NHTBH38 test set. NA means the data is not available because
            of the SCF convergence issue.}
    \label{tab:RB_NHTBH38}
    \scalebox{0.9}{
    \begin{tabular}{@{}lr|rrr|rrr|rrr@{}}
    \toprule
        &  & \multicolumn{3}{c}{DFA}
        & \multicolumn{3}{c}{post-LOSC2}
        & \multicolumn{3}{c}{SCF-LOSC2} \\ \hline
    case & Ref & LDA    & BLYP  & B3LYP
    & LDA    & BLYP          & B3LYP
    & LDA    & BLYP         & B3LYP \\ \midrule
\multicolumn{11}{l}{Forward: $E_{\rm{TS}}$ - $\sum E_{\rm{reactant}}$} \\ \hline
    TS1  & 17.13  & 2.47   & 8.51   & 11.34 & 2.48           & 8.51            & 11.34            & 2.48          & 8.51           & 11.34           \\
    TS2  & 42.18  & 18.24  & 25.97  & 30.96 & 18.24          & 25.96           & 30.96            & 18.24         & 25.96          & 30.96           \\
    TS3  & 18     & 2.18   & 9.86   & 12.47 & 2.18           & 9.86            & 12.47            & 2.18          & 9.86           & 12.47           \\
    TS4  & 30.38  & 13.2   & 16.08  & 21.75 & 13.21          & 16.08           & 21.75            & 13.21         & 16.08          & 21.75           \\
    TS5  & 2.27   & -15.83 & -11.46 & -7.31 & -15.83         & -11.46          & -7.31            & -15.83        & -11.46         & -7.31           \\
\bottomrule
    \end{tabular}
    }
\end{table}

\clearpage
\newpage
\renewcommand\arraystretch{0.6}
\begin{longtable}{lr|rrr|rrr|rrr}
    \caption{Detailed data for the ionization potentials (in eV) of the
            IP test set. NA means the data is not available because
            of the SCF convergence issue.}
    \label{tab:IP}\\
    \toprule
        &  & \multicolumn{3}{c}{DFA}
        & \multicolumn{3}{c}{post-LOSC2}
        & \multicolumn{3}{c}{SCF-LOSC2} \\ \hline
    Molecule & Ref & LDA    & BLYP  & B3LYP
    & LDA    & BLYP          & B3LYP
    & LDA    & BLYP         & B3LYP \\ \midrule
    \endfirsthead
    \multicolumn{11}{l}{Table \ref{tab:IP} continued.} \\
    \toprule
        &  & \multicolumn{3}{c}{DFA}
        & \multicolumn{3}{c}{post-LOSC2}
        & \multicolumn{3}{c}{SCF-LOSC2} \\ \hline
    Molecule & Ref & LDA    & BLYP  & B3LYP
    & LDA    & BLYP          & B3LYP
    & LDA    & BLYP         & B3LYP \\ \midrule
    \endhead
    \midrule
    \endfoot
    \endlastfoot
    H4C          & 14.4  & 9.45  & 9.37  & 10.75 & 13.82          & 13.75           & 14.26            & 13.81         & 13.74          & 14.25           \\
    H3N          & 11.03 & 6.33  & 6.17  & 7.56  & 11.54          & 11.4            & 11.75            & 11.54         & 11.39          & 11.75           \\
    HO           & 13.07 & 7.4   & 7.35  & 8.99  & 13.55          & 13.54           & 13.96            & 13.55         & 13.53          & 13.95           \\
    H2O          & 12.74 & 7.37  & 7.18  & 8.81  & 13.4           & 13.24           & 13.67            & 13.4          & 13.23          & 13.67           \\
    HF           & 16.2  & 9.78  & 9.59  & 11.51 & 16.98          & 16.8            & 17.3             & 16.97         & 16.79          & 17.3            \\
    H4Si         & 12.84 & 8.51  & 8.43  & 9.65  & 12.02          & 11.99           & 12.51            & 12.01         & 11.97          & 12.51           \\
    HP           & 10.18 & 6.15  & 5.89  & 7.02  & 9.59           & 9.33            & 9.78             & 9.59          & 9.33           & 9.78            \\
    H2P          & 9.82  & 5.98  & 5.78  & 6.89  & 9.4            & 9.21            & 9.63             & 9.4           & 9.21           & 9.63            \\
\bottomrule
\end{longtable}

\clearpage
\newpage
\renewcommand\arraystretch{0.6}
\begin{longtable}{lr|rrr|rrr|rrr}
    \caption{Detailed data for the electron affinities (in eV) of the
            EA test set. NA means the data is not available because
            of the SCF convergence issue.}
    \label{tab:EA}\\
    \toprule
        &  & \multicolumn{3}{c}{DFA}
        & \multicolumn{3}{c}{post-LOSC2}
        & \multicolumn{3}{c}{SCF-LOSC2} \\ \hline
    Molecule & Ref & LDA    & BLYP  & B3LYP
    & LDA    & BLYP          & B3LYP
    & LDA    & BLYP         & B3LYP \\ \midrule
    \endfirsthead
    \multicolumn{11}{l}{Table \ref{tab:EA} continued.} \\
    \toprule
        &  & \multicolumn{3}{c}{DFA}
        & \multicolumn{3}{c}{post-LOSC2}
        & \multicolumn{3}{c}{SCF-LOSC2} \\ \hline
    Molecule & Ref & LDA    & BLYP  & B3LYP
    & LDA    & BLYP          & B3LYP
    & LDA    & BLYP         & B3LYP \\ \midrule
    \endhead
    \midrule
    \endfoot
    \endlastfoot
    HC     & 1.19  & 5.78 & 5.16 & 4.15  & 1.16           & 0.66            & 0.73             & 1.16          & 0.66           & 0.73            \\
    H2C    & 0.83  & 3.9  & 3.6  & 2.69  & -0.35          & -0.59           & -0.46            & -0.36         & -0.59          & -0.46           \\
    H3C    & -0.06 & 3.59 & 3.06 & 2.2   & -0.45          & -0.84           & -0.72            & -0.45         & -0.84          & -0.73           \\
    HN     & 0.33  & 4.86 & 4.5  & 3.3   & -0.14          & -0.41           & -0.34            & -0.14         & -0.41          & -0.34           \\
    H2N    & 0.75  & 5.22 & 4.62 & 3.45  & 0.23           & -0.23           & -0.17            & 0.23          & -0.24          & -0.18           \\
    HO     & 1.83  & 7.24 & 6.53 & 5.03  & 1.14           & 0.59            & 0.55             & 1.14          & 0.58           & 0.55            \\
    HSi    & 1.26  & 4.44 & NA   & 3.44  & 1.47           & NA              & 1.16             & 1.46          & NA             & 1.15            \\
    H2Si   & 1.07  & 4.25 & 3.91 & 3.34  & 1.27           & 0.96            & 1.03             & 1.27          & 0.96           & 1.03            \\
\bottomrule
\end{longtable}

\clearpage
\newpage

\section{II. macro-SCF-LOSC}
\begin{figure}[htbp]
    \centering
    \includegraphics[width=0.6\linewidth]{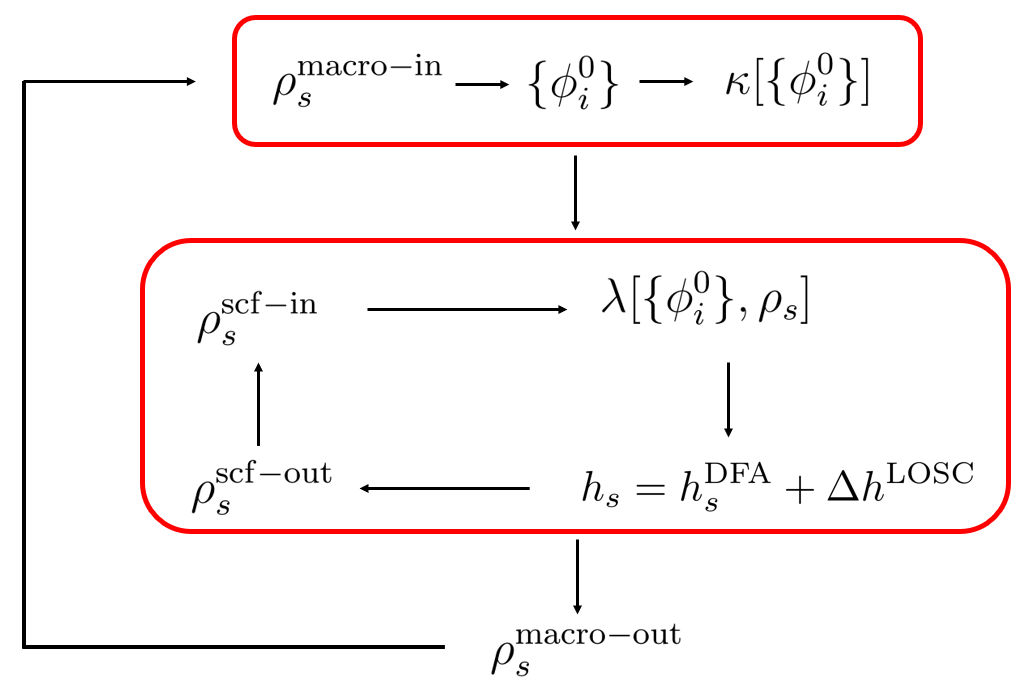}
    \caption{The macro-SCF procedure for LOSC.}
    \label{fig:macro_scf}
\end{figure}
The working flow of the macro-SCF-LOSC is shown in Figure \ref{fig:macro_scf}.
The macro-SCF-LOSC includes two layers of SCF cycle. The outer lay SCF cycle is to
update the LOs. The inner lay SCF cycle is to update
the electron density with the given LOs. The inner lay SCF cycle
shares the same spirit of the new SCF-LOSC approach.
In details, the macro-SCF-LOSC involves the following steps:
(1) start the macro-SCF-LOSC with an initial density guess
$\rho_s^{\rm{macro-in}}$;
(2) apply LOSC localization procedure to produce LOs $\{\phi_i^0\}$
based on $\rho_s^{\rm{macro-in}}$;
(3) construct LOSC curvature matrix from $\{\phi_i^0\}$;
(4) achieve the convergence of the inner SCF-LOSC calculation with
the given $\{\phi_i^0\}$ and $\kappa[\{\phi_i^0\}]$ and
obtain the converged density as $\rho_s^{\rm{macro-out}}$;
(5)check the convergence of the macro-SCF-LOSC; go back to step
(2) if it is not converged.

\renewcommand\arraystretch{1.1}
\begin{table}[]
    \caption{Total energies (in a.u.) comparison between SCF-LOSC2 and
        macro-SCF-LOSC2.
        LOSC2 calculation is associated with B3LYP functional.
        LiH and LiF molecules are stretched at 14 bohr,
        and HF molecule is stretched at 16 bohr.}
    \label{tab:E_tot_scf_vs_macro}
    \begin{tabular}{@{}llll@{}}
    \toprule
    Molecule & SCF-LOSC2     & macro-SCF-LOSC2 & $\Delta E$\footnotemark[1]      \\ \midrule
    HF  & -100.2655352 & -100.2655352   & -1.01E-12 \\
    LiH & -7.994964751 & -7.994964802   & -5.06E-08 \\
    LiF & -107.2563249 & -107.2565187   & -1.94E-04 \\ \bottomrule
    \end{tabular}\\
    \footnotemark[1] {The energy difference between SCF-LOSC2 and
    macro-SCF-LOSC2.}
\end{table}

\renewcommand\arraystretch{1.1}
\begin{table}[]
    \caption{Mulliken charge (in a.u.) comparison between SCF-LOSC2 and
        macro-SCF-LOSC2.
        LOSC2 calculation is associated with B3LYP functional.
        LiH and LiF molecules are stretched at 14 bohr,
        and HF molecule is stretched at 16 bohr. The Mulliken charge
        refers to F atom in HF, H atom in LiH and F atom in LiF.}
    \label{tab:mk_scf_vs_macro}
    \begin{tabular}{@{}llll@{}}
    \toprule
    Molecule & SCF-LOSC     & macro-SCF-LOSC & Diff (macro - scf)      \\ \midrule
    HF  & 0.000000 & 0.000000       & 0.00E+00  \\
    LiH & 0.000313 & 0.000311       & -2.00E-06 \\
    LiF & 0.000310 & 0.000240       & -7.00E-05 \\ \bottomrule
    \end{tabular}
\end{table}

To compare the performance between new SCF-LOSC (shown in Figure 2 of
the main text) and the macro-SCF-LOSC,
Table \ref{tab:E_tot_scf_vs_macro} and \ref{tab:mk_scf_vs_macro} show
the total energies and Mulliken charges from these two methods
for stretched HF, LiH and LiF molecules at large distance. In the cases of stretched
HF and LiH, in which the parent
functional B3LYP already describes the electron density well,
the results from SCF-LOSC and macro-SCF-LOSC are almost the same.
In the case of stretched LiF, in which the
parent functional B3LYP yields delocalized electron density,
SCF-LOSC still provide similar
results to the ones from macro-SCF-LOSC with the difference of total
energy only up to 1e-4 a.u. ($< 0.1$ kcal/mol). Those results indicate
the results from the new SCF-LOSC is reliable. Also it suggests that
applying macro-SCF-LOSC would not be necessary in practice.

\bibliography{qm4d,FL_LOSC}